\documentclass[11pt,a4paper,english]{article}
\usepackage[T1]{fontenc}
\usepackage[latin1]{inputenc}
\usepackage{subfigure}
\usepackage{amsmath}
\usepackage{graphicx}
\usepackage{amssymb}

\makeatletter

\providecommand{\tabularnewline}{\\}

\usepackage{a4wide}
\usepackage{psfrag}
\makeatletter

\@addtoreset{equation}{section}

\@addtoreset{figure}{section}

\@addtoreset{table}{section}
\makeatother 

\usepackage{babel}
\makeatother
\begin{document}

\title{\begin{flushright}{\normalsize ITP-Budapest Report No. 634}\end{flushright}\vspace{1cm}Form
factors in finite volume I:\\
form factor bootstrap and truncated conformal space}

\author{B. Pozsgay$^{1}$%
\thanks{E-mail: pozsi@bolyai.elte.hu%
} ~and G. Takács$^{2}$%
\thanks{E-mail: takacs@elte.hu%
}\\
\\
$^{1}$\emph{Institute for Theoretical Physics}\\
\emph{Eötvös University, Budapest}\\
\emph{}\\
$^{2}$\emph{HAS Research Group for Theoretical Physics}\\
\emph{H-1117 Budapest, Pázmány Péter sétány 1/A}}

\date{26th June 2007}

\maketitle
\begin{abstract}
We describe the volume dependence of matrix elements of local fields
to all orders in inverse powers of the volume (i.e. only neglecting
contributions that decay exponentially with volume). Using the scaling
Lee-Yang model and the Ising model in a magnetic field as testing
ground, we compare them to matrix elements extracted in finite volume
using truncated conformal space approach to exact form factors obtained
using the bootstrap method. We obtain solid confirmation for the form
factor bootstrap, which is different from all previously available
tests in that it is a non-perturbative and direct comparison of exact
form factors to multi-particle matrix elements of local operators,
computed from the Hamiltonian formulation of the quantum field theory.
We also demonstrate that combining form factor bootstrap and truncated
conformal space is an effective method for evaluating finite volume
form factors in integrable field theories over the whole range in
volume.
\end{abstract}

\section{Introduction}

The matrix elements of local operators (form factors) are central
objects in quantum field theory. In two-dimensional integrable quantum
field theory the $S$ matrix can be obtained exactly in the framework
of factorized scattering (see \cite{ZZ79,Sbstr} for reviews), and
using the scattering amplitudes as input it is possible to obtain
a set of axioms satisfied by the form factors \cite{karowski}, which
provides the basis for the so-called form factor bootstrap (see \cite{Smirnov}
for a review). 

Although the connection with the Lagrangian formulation of quantum
field theory is rather indirect in the bootstrap approach, it is thought
that the general solution of the form factor axioms determines the
complete local operator algebra of the theory \cite{cardy_mussardo}.
This expectation was confirmed in many cases by explicit comparison
of the space of solutions to the spectrum of local operators as described
by the ultraviolet limiting conformal field theory \cite{koubek_mussardo,koubek1,koubek2,smirnovcounting};
mathematical foundation is provided by the local commutativity theorem
stating that operators specified by solutions of the form factor bootstrap
are mutually local \cite{Smirnov}. Another important piece of information
comes from correlation functions: using form factors, a spectral representation
for the correlation functions can be built which provides a large
distance expansion \cite{isingspincorr,Z1}, while the Lagrangian
or perturbed conformal field theory formulation allows one to obtain
a short-distance expansion, which can then be compared provided there
is an overlap between their regimes of validity \cite{Z1}. Other
evidence for the correspondence between the field theory and the solutions
of the form factor bootstrap results from evaluating sum rules like
Zamolodchikov's $c$-theorem \cite{c-th,c-thspectral} or the $\Delta$-theorem
\cite{delta-th}, both of which can be used to express conformal data
as spectral sums in terms of form factors. Direct comparisons with
multi-particle matrix elements are not so readily available, except
for perturbative or $1/N$ calculations in some simple cases \cite{karowski}. 

Therefore, part of the motivation of this paper is to provide non-perturbative
evaluation of form factors from the Hamiltonian formulation, which
then allows for a direct comparison with solutions of the form factor
axioms. Another goal is to have a better understanding of finite size
effects in the case of matrix elements of local operators, and to
contribute to the investigation of finite volume \cite{finitevolFF}
(and also finite temperature \cite{Doyon}) form factors and correlation
functions.

Based on what we learned from our previous investigation of decay
rates in finite volume \cite{takacspozsgay}, in this paper we determine
form factors using a formulation of quantum field theory in finite
volume. In two space-time dimensions this is most efficiently done
using the truncated conformal space approach (TCSA) developed by Yurov
and Zamolodchikov \cite{yurov_zamolodchikov}, but one could have
also made use of, for example, lattice field theory: the only important
point is to have a method which can be used to determine energy levels
and matrix elements of local operators as functions of the volume.
We give a relation between finite and infinite volume multi-particle
form factors, which is a natural extension of the results by Lellouch
and Lüscher for two-particle decay matrix elements \cite{lellouch}.
It is used to determine form factors in two important examples of
integrable quantum field theory: the scaling Lee-Yang model and the
Ising model in a magnetic field. We show that these results agree
very well with the predictions of the form factor bootstrap. We only
treat matrix elements of local fields between multi-particle states
for which there are no disconnected pieces (which appear whenever
there are particles with coincident rapidities in the left and right
multi-particle states); the treatment of disconnected pieces, together
with a number of theoretical arguments and ramifications, are postponed
to a subsequent publication \cite{crossing}.

The organization of the paper is as follows. In section 2, after a
brief review of necessary facts concerning the form factor bootstrap
we give the general description of form factors (without disconnected
pieces) to all orders in $1/L$ based on an analysis of two-point
correlation functions. Section 3 describes the two models we chose
for demonstration, and also specifies the method of evaluating form
factors from truncated conformal space. Numerical results on elementary
form factors (vacuum--many-particle matrix elements) are given in
section 4, while the general case is treated in section 5. We give
our conclusions in section 6.

\section{Form factors in finite volume}

\subsection{Form factor bootstrap}

Here we give a very brief summary of the axioms of the form factor
bootstrap, in order to set up notations and to provide background
for later arguments; the interested reader is referred to Smirnov's
review \cite{Smirnov} for more details. Let us suppose for simplicity
that the spectrum of the model consists of particles $A_{i}$, $i=1,\dots,N$
with masses $m_{i}$, which are assumed to be strictly non-degenerate,
i.e. $m_{i}\neq m_{j}$ for any $i\neq j$ (and therefore also self-conjugate).
Because of integrability, multi-particle scattering amplitudes factorize
into the product of pairwise two-particle scatterings, which are purely
elastic (in other words: diagonal). This means that any two-particle
scattering amplitude is a pure phase, which we denote by $S_{ij}\left(\theta\right)$
where $\theta$ is the relative rapidity of the incoming particles
$A_{i}$ and $A_{j}$. Incoming and outgoing asymptotic states can
be distinguished by ordering of the rapidities:\[
|\theta_{1},\dots,\theta_{n}\rangle_{i_{1}\dots i_{n}}=\begin{cases}
|\theta_{1},\dots,\theta_{n}\rangle_{i_{1}\dots i_{n}}^{in} & :\;\theta_{1}>\theta_{2}>\dots>\theta_{n}\\
|\theta_{1},\dots,\theta_{n}\rangle_{i_{1}\dots i_{n}}^{out} & :\;\theta_{1}<\theta_{2}<\dots<\theta_{n}\end{cases}\]
and states which only differ in the order of rapidities are related
by\[
|\theta_{1},\dots,\theta_{k},\theta_{k+1},\dots,\theta_{n}\rangle_{i_{1}\dots i_{k}i_{k+1}\dots i_{n}}=S_{i_{k}i_{k+1}}(\theta_{k}-\theta_{k+1})|\theta_{1},\dots,\theta_{k+1},\theta_{k},\dots,\theta_{n}\rangle_{i_{1}\dots i_{k+1}i_{k}\dots i_{n}}\]
from which the $S$ matrix of any multi-particle scattering process
can be obtained. The normalization of these states is specified by
the following inner product for the one-particle states:\[
\,_{j}\langle\theta^{'}|\theta\rangle_{i}=\delta_{ij}2\pi\delta(\theta^{'}-\theta)\]
The form factors of a local operator $\mathcal{O}(t,x)$ are defined
as\begin{equation}
F_{mn}^{\mathcal{O}}(\theta_{1}^{'},\dots,\theta_{m}^{'}|\theta_{1},\dots,\theta_{n})_{j_{1}\dots j_{m};i_{1}\dots i_{n}}=\,_{j_{1}\dots j_{m}}\langle\theta_{1}^{'},\dots,\theta_{m}^{'}\vert\mathcal{O}(0,0)\vert\theta_{1},\dots,\theta_{n}\rangle_{i_{1}\dots i_{n}}\label{eq:genff}\end{equation}
With the help of the crossing relations\begin{eqnarray}
 &  & F_{mn}^{\mathcal{O}}(\theta_{1}^{'},\dots,\theta_{m}^{'}|\theta_{1},\dots,\theta_{n})_{j_{1}\dots j_{m};i_{1}\dots i_{n}}=\nonumber \\
 &  & \qquad F_{m-1n+1}^{\mathcal{O}}(\theta_{1}^{'},\dots,\theta_{m-1}^{'}|\theta_{m}^{'}+i\pi,\theta_{1},\dots,\theta_{n})_{j_{1}\dots j_{m-1};j_{m}i_{1}\dots i_{n}}\nonumber \\
 &  & \qquad+\sum_{k=1}^{n}\Big(2\pi\delta_{j_{m}i_{k}}\delta(\theta_{m}^{'}-\theta_{k})\prod_{l=1}^{k-1}S_{i_{l}i_{k}}(\theta_{l}-\theta_{k})\times\nonumber \\
 &  & \qquad F_{m-1n-1}^{\mathcal{O}}(\theta_{1}^{'},\dots,\theta_{m-1}^{'}|\theta_{1},\dots,\theta_{k-1},\theta_{k+1}\dots,\theta_{n})_{j_{1}\dots j_{m-1};j_{m}i_{1}\dots i_{k-1}i_{k+1}\dots i_{n}}\Big)\label{eq:ffcrossing}\end{eqnarray}
all form factors can be expressed in terms of the elementary form
factors\[
F_{n}^{\mathcal{O}}(\theta_{1},\dots,\theta_{n})_{i_{1}\dots i_{n}}=\langle0\vert\mathcal{O}(0,0)\vert\theta_{1},\dots,\theta_{n}\rangle_{i_{1}\dots i_{n}}\]
which satisfy the following axioms: 

I. Exchange:

\begin{center}\begin{eqnarray}
 &  & F_{n}^{\mathcal{O}}(\theta_{1},\dots,\theta_{k},\theta_{k+1},\dots,\theta_{n})_{i_{1}\dots i_{k}i_{k+1}\dots i_{n}}=\nonumber \\
 &  & \qquad S_{i_{k}i_{k+1}}(\theta_{k}-\theta_{k+1})F_{n}^{\mathcal{O}}(\theta_{1},\dots,\theta_{k+1},\theta_{k},\dots,\theta_{n})_{i_{1}\dots i_{k+1}i_{k}\dots i_{n}}\label{eq:exchangeaxiom}\end{eqnarray}
\par\end{center}

II. Cyclic permutation: \begin{equation}
F_{n}^{\mathcal{O}}(\theta_{1}+2i\pi,\theta_{2},\dots,\theta_{n})=F_{n}^{\mathcal{O}}(\theta_{2},\dots,\theta_{n},\theta_{1})\label{eq:cyclicaxiom}\end{equation}

III. Kinematical singularity\begin{equation}
-i\mathop{\textrm{Res}}_{\theta=\theta^{'}}F_{n+2}^{\mathcal{O}}(\theta+i\pi,\theta^{'},\theta_{1},\dots,\theta_{n})_{i\, j\, i_{1}\dots i_{n}}=\left(1-\delta_{i\, j}\prod_{k=1}^{n}S_{i\, i_{k}}(\theta-\theta_{k})\right)F_{n}^{\mathcal{O}}(\theta_{1},\dots,\theta_{n})_{i_{1}\dots i_{n}}\label{eq:kinematicalaxiom}\end{equation}

IV. Dynamical singularity \begin{equation}
-i\mathop{\textrm{Res}}_{\theta=\theta^{'}}F_{n+2}^{\mathcal{O}}(\theta+i\bar{u}_{jk}^{i}/2,\theta^{'}-i\bar{u}_{ik}^{j}/2,\theta_{1},\dots,\theta_{n})_{i\, j\, i_{1}\dots i_{n}}=\Gamma_{ij}^{k}F_{n+1}^{\mathcal{O}}(\theta,\theta_{1},\dots,\theta_{n})_{k\, i_{1}\dots i_{n}}\label{eq:dynamicalaxiom}\end{equation}
whenever $k$ occurs as the bound state of the particles $i$ and
$j$, corresponding to a bound state pole of the $S$ matrix of the
form\begin{equation}
S_{ij}(\theta\sim iu_{ij}^{k})\sim\frac{i\left(\Gamma_{ij}^{k}\right)^{2}}{\theta-iu_{ij}^{k}}\label{eq:Smatpole}\end{equation}
where $\Gamma_{ij}^{k}$ is the on-shell three-particle coupling and
$u_{ij}^{k}$ is the so-called fusion angle. The fusion angles satisfy\begin{eqnarray*}
m_{k}^{2} & = & m_{i}^{2}+m_{j}^{2}+2m_{i}m_{j}\cos u_{ij}^{k}\\
2\pi & = & u_{ij}^{k}+u_{ik}^{j}+u_{jk}^{i}\end{eqnarray*}
and we also used the notation $\bar{u}_{ij}^{k}=\pi-u_{ij}^{k}$.
Axioms I-IV are supplemented by the assumption of maximum analyticity
(i.e. that the form factors are meromorphic functions which only have
the singularities prescribed by the axioms) and possible further conditions
expressing properties of the particular operator whose form factors
are sought.

\subsection{Finite size corrections for form factors}

Let us consider the spectral representation of the Euclidean two-point
function\begin{eqnarray}
\langle\mathcal{O}(\bar{x})\mathcal{O}'(0,0)\rangle & = & \sum_{n=0}^{\infty}\sum_{i_{1}\dots i_{n}}\left(\prod_{k=1}^{n}\int_{-\infty}^{\infty}\frac{d\theta_{k}}{2\pi}\right)F_{n}^{\mathcal{O}}(\theta_{1},\theta_{2},\dots,\theta_{n})_{i_{1}\dots i_{n}}\times\nonumber \\
 &  & \quad F_{n}^{\mathcal{O}'}(\theta_{1},\theta_{2},\dots,\theta_{n})_{i_{1}\dots i_{n}}^{+}\exp\left(-r\sum_{k=1}^{n}m_{i_{k}}\cosh\theta_{k}\right)\label{eq:spectralrepr}\end{eqnarray}
where\[
F_{n}^{\mathcal{O}'}(\theta_{1},\theta_{2},\dots,\theta_{n})_{i_{1}\dots i_{n}}^{+}=\,_{i_{1}\dots i_{n}}\langle\theta_{1},\dots,\theta_{n}|\mathcal{O}'(0,0)|0\rangle=F_{n}^{\mathcal{O}'}(\theta_{1}+i\pi,\theta_{2}+i\pi,\dots,\theta_{n}+i\pi)_{i_{1}\dots i_{n}}\]
(which is just the complex conjugate of $F_{n}^{\mathcal{O}'}$ for
unitary theories) and $r=\sqrt{\tau^{2}+x^{2}}$ is the length of
the Euclidean separation vector $\bar{x}=(\tau,x)$. 

In finite volume $L$, the space of states can still be labeled by
multi-particle states but the momenta (and therefore the rapidities)
are quantized. Denoting the quantum numbers $I_{1},\dots,I_{n}$ the
two-point function of the same local operator can be written as\begin{eqnarray}
\langle\mathcal{O}(\tau,0)\mathcal{O}'(0,0)\rangle_{L} & = & \sum_{n=0}^{\infty}\sum_{i_{1}\dots i_{n}}\sum_{I_{1}\dots I_{n}}\langle0\vert\mathcal{O}(0,0)\vert\{ I_{1},I_{2},\dots,I_{n}\}\rangle_{i_{1}\dots i_{n},L}\times\nonumber \\
 &  & \quad_{i_{1}\dots i_{n}}\langle\{ I_{1},I_{2},\dots,I_{n}\}|\mathcal{O}'(0,0)|0\rangle_{L}\exp\left(-\tau\sum_{k=1}^{n}m_{i_{k}}\cosh\theta_{k}\right)\label{eq:finitevolspectralrepr}\end{eqnarray}
where we supposed that the finite volume multi-particle states $\vert\{ I_{1},I_{2},\dots,I_{n}\}\rangle_{i_{1}\dots i_{n},L}$
are orthonormal and for simplicity restricted the formula to separation
in Euclidean time $\tau$ only. The index $L$ signals that the matrix
element is evaluated in finite volume $L$. Using the finite volume
expansion developed by Lüscher in \cite{luscher_onept} one can easily
see that \begin{equation}
\langle\mathcal{O}(\tau,0)\mathcal{O}'(0)\rangle-\langle\mathcal{O}(\tau,0)\mathcal{O}'(0)\rangle_{L}\sim O(\mathrm{e}^{-\mu L})\label{eq:corrfinvoldiff}\end{equation}
where $\mu$ is some characteristic mass scale. 

Note that Lüscher's finite volume expansion is derived using covariant
perturbation theory, but the final expansion is obtained after resummation
in the coupling constant, and is expected to hold non-perturbatively.
The finite volume corrections are then expressed in terms of the exact
one-particle-irreducible vertex functions of the infinite volume theory.
According to Lüscher's classification of finite volume Feynman graphs,
the difference between the finite and infinite volume correlation
function is given by contributions from graphs of nontrivial gauge
class, i.e. graphs in which some propagator has a nonzero winding
number around the cylinder. Such graphs always carry an exponential
suppression factor in $L$, whose exponent can be determined by analyzing
the singularities of the propagators and vertex functions entering
the expressions. In a massive theory, all such singularities lie away
from the real axis of the Mandelstam variables, and the one with the
smallest imaginary part determines $\mu$. It turns out that the value
of $\mu$ is determined by the exact mass spectrum of the particles
and also the bound state fusions between them \cite{luscher_onept,klassenmelzer}.
Therefore it is universal, which means that it is independent of the
correlation function considered. In general $\mu\leq m_{1}$ where
$m_{1}$ is the lightest particle mass (the mass gap of the theory),
because there are always corrections in which the lightest particle
loops around the finite volume $L$, and so the mass shell pole of
the corresponding exact propagator is always present. Contributions
from such particle loops to the vacuum expectation value are evaluated
in subsection 4.1 (for a graphical representation see figure \ref{fig:vevLcorr}),
while an example of a finite volume correction corresponding to a
bound state fusion (a so-called $\mu$-term) is discussed in subsection
4.1.2 (figure \ref{fig:ising3pcorr}).

To relate the finite and infinite volume form factors a further step
is necessary, because the integrals in the spectral representation
(\ref{eq:spectralrepr}) must also be discretized. Let us consider
this problem first for the case of free particles:\[
\left(\prod_{k=1}^{n}\int_{-\infty}^{\infty}\frac{d\theta_{k}}{2\pi}\right)f(\theta_{1},\dots,\theta_{n})=\left(\prod_{k=1}^{n}\int_{-\infty}^{\infty}\frac{dp_{k}}{2\pi\omega_{k}}\right)f(p_{1},\dots,p_{n})\]
where \[
p_{k}=m_{i_{k}}\sinh\theta_{k}\quad,\quad\omega_{k}=m_{i_{k}}\cosh\theta_{k}\]
are the momenta and energies of the particles. In finite volume\[
p_{k}=\frac{2\pi I_{k}}{L}\]
and it is well-known (as a consequence of the Poisson summation formula,
cf. \cite{luscher_2particle}) that\begin{equation}
\sum_{I_{1},\dots I_{n}}g\left(\frac{2\pi I_{1}}{L},\dots,\frac{2\pi I_{n}}{L}\right)=\left(\frac{L}{2\pi}\right)^{n}\left(\prod_{k=1}^{n}\int_{-\infty}^{\infty}dp_{k}\right)g(p_{1},\dots,p_{n})+O(L^{-N})\label{eq:intsumcorr}\end{equation}
provided the function $g$ and its first $N$ derivatives are integrable.
Recalling that form factors are analytic functions for real momenta,
in our case this is true for derivatives of any order, due to the
exponential suppression factor in the spectral integrals, provided
the form factors grow at most polynomially in the momentum, i.e.\[
\left|F_{n}(\theta_{1}+\theta,\theta_{2}+\theta,\dots,\theta_{n}+\theta)\right|\sim\mathrm{e}^{x\left|\theta\right|}\quad\mathrm{as}\quad\left|\theta\right|\rightarrow\infty\]
This is true if we only consider operators which have a power-like
short distance singularity in their two-point functions \cite{delfino_mussardo}:\[
\langle0\vert\mathcal{O}(\bar{x})\mathcal{O}(0)\vert0\rangle=\frac{1}{r^{2\Delta}}\]
Such operators are called scaling fields and are generally assumed
to form a closed algebra under the operator product expansion. 

Therefore the discrete sum differs from the continuum integral only
by terms decaying faster than any power in $1/L$, i.e. by terms exponentially
suppressed in $L$. Taking into account that (\ref{eq:spectralrepr})
is valid for any pair $\mathcal{O}$, $\mathcal{O}'$ of scaling fields,
we obtain \begin{equation}
\langle0\vert\mathcal{O}(0,0)\vert\{ I_{1},\dots,I_{n}\}\rangle_{i_{1}\dots i_{n},L}=\frac{1}{\sqrt{\rho_{i_{1}\dots i_{n}}^{(0)}(\tilde{\theta}_{1},\dots,\tilde{\theta}_{n})}}F_{n}^{\mathcal{O}}(\tilde{\theta}_{1},\dots,\tilde{\theta}_{n})_{i_{1}\dots i_{n}}+O(\mathrm{e}^{-\mu'L})\label{eq:freepartffrelation}\end{equation}
where\[
\sinh\tilde{\theta}_{k}=\frac{2\pi I_{k}}{m_{i_{k}}L}\]
and\begin{equation}
\rho_{i_{1}\dots i_{n}}^{(0)}(\tilde{\theta}_{1},\dots,\tilde{\theta}_{n})=\prod_{k=1}^{n}m_{i_{k}}L\cosh\tilde{\theta}_{k}\label{eq:freejacobidet}\end{equation}
$\rho_{n}^{(0)}$ is nothing else than the Jacobi determinant corresponding
to changing from the variables $2\pi I_{k}$ to the rapidities $\tilde{\theta}_{k}$.
The term $O(\mathrm{e}^{-\mu'L})$ signifies that our considerations
are valid to all orders in $1/L$, although our argument does not
tell us the value of $\mu'$: to do that, we would need more information
about the correction term in the discretization (\ref{eq:intsumcorr}). 

In the case of interacting particles a more careful analysis is necessary
because the quantization rules are different from the free case. In
a two-dimensional integrable quantum field theory, general multi-particle
levels are determined by the Bethe-Yang equations\begin{equation}
Q_{k}(\tilde{\theta}_{1},\dots,\tilde{\theta}_{n})_{i_{1}\dots i_{n}}=m_{i_{k}}L\sinh\tilde{\theta}_{k}+\sum_{l\neq k}\delta_{i_{k}i_{l}}(\tilde{\theta}_{k}-\tilde{\theta}_{l})=2\pi I_{k}\quad,\quad k=1,\dots,n\label{eq:betheyang}\end{equation}
which are valid to any order in $1/L$, where\begin{equation}
\delta_{ij}(\theta)=-i\log S_{ij}(\theta)\label{eq:origphaseshift}\end{equation}
are the two-particle scattering phase-shifts. Therefore the proper
generalization of (\ref{eq:freepartffrelation}) is\begin{equation}
\langle0\vert\mathcal{O}(0,0)\vert\{ I_{1},\dots,I_{n}\}\rangle_{i_{1}\dots i_{n},L}=\frac{1}{\sqrt{\rho_{i_{1}\dots i_{n}}(\tilde{\theta}_{1},\dots,\tilde{\theta}_{n})}}F_{n}^{\mathcal{O}}(\tilde{\theta}_{1},\dots,\tilde{\theta}_{n})_{i_{1}\dots i_{n}}+O(\mathrm{e}^{-\mu'L})\label{eq:ffrelation}\end{equation}
where\begin{eqnarray}
\rho_{i_{1}\dots i_{n}}(\theta_{1},\dots,\theta_{n}) & = & \det\mathcal{J}^{(n)}(\theta_{1},\dots,\theta_{n})_{i_{1}\dots i_{n}}\label{eq:byjacobian}\\
 &  & \mathcal{J}_{kl}^{(n)}(\theta_{1},\dots,\theta_{n})_{i_{1}\dots i_{n}}=\frac{\partial Q_{k}(\theta_{1},\dots,\theta_{n})_{i_{1}\dots i_{n}}}{\partial\theta_{l}}\quad,\quad k,l=1,\dots,n\nonumber \end{eqnarray}
and $\tilde{\theta}_{k}$ are the solutions of the Bethe-Yang equations
(\ref{eq:betheyang}) corresponding to the state with the specified
quantum numbers $I_{1},\dots,I_{n}$ at the given volume $L$.

Similar arguments were previously used to obtain the finite size dependence
of kaon decay matrix elements by Lin et al. \cite{sachrajda}; they
also give a more detailed analysis of the discretization in (\ref{eq:intsumcorr}).
In principle it could be that $\mu'<\mu$ if the discretization introduces
larger errors than the finite size correction of the two-point functions.
However, it is possible to give an argument that actually $\mu'=\mu$.
Recall that the Poisson formula gives the discrete sum in terms of
a Fourier transform: the leading term is the Fourier transform of
the summand evaluated at wave number $0$ (i.e. the integral) and
the corrections are determined by the decay of the Fourier transform
at large wave numbers. The function we need to consider is\begin{eqnarray}
h(p_{1},\dots,p_{n}) & = & \sum_{i_{1}\dots i_{n}}\rho_{i_{1}\dots i_{n}}(\theta_{1},\dots,\theta_{n})^{-1}F_{n}^{\mathcal{O}}(\theta_{1},\theta_{2},\dots,\theta_{n})_{i_{1}\dots i_{n}}F_{n}^{\mathcal{O}'}(\theta_{1},\theta_{2},\dots,\theta_{n})_{i_{1}\dots i_{n}}^{+}\times\nonumber \\
 &  & \exp\left(-r\sum_{k=1}^{n}m_{i_{k}}\cosh\theta_{k}\right)\label{eq:hfunction}\end{eqnarray}
where $p_{k}=m_{i_{k}}\sinh\theta_{k}$ are the momentum variables.
Due to the analyticity of the form factors for real rapidities, this
function is analytic for physical (real) momenta, and together with
all of its derivatives decays more rapidly than any power at infinity.
Therefore its Fourier transform taken in the momentum variables has
the same asymptotic property, i.e. it (and its derivatives) decay
more rapidly than any power:\[
\tilde{h}(\kappa_{1},\dots,\kappa_{n})\sim\mathrm{e}^{-\mu''|\kappa|}\]
for large $\kappa$. As a result, discretization introduces an error
of order $\mathrm{e}^{-\mu''L}$. The asymptotic exponent $\mu''$
of the Fourier transform can be generally determined by shifting the
contour of the integral transform and is given by the position of
the singularity closest to the real momentum domain (this is essentially
the procedure that Lüscher uses in \cite{luscher_onept}). Singularities
of the form factors are given by the same analytic structure as that
of the amplitudes which determine the exponent $\mu$ in eqn. (\ref{eq:corrfinvoldiff})
(see the discussion there). Thus we expect that $\mu''=\mu$ and,
as a result, $\mu'=\mu$. 

This argument is just an intuitive reasoning, although it can be made
a little more precise. First of all, we must examine whether the determinant
$\rho_{i_{1}\dots i_{n}}(\theta_{1},\dots,\theta_{n})$ can have any
zeros. It can always be written in the form\[
\rho_{i_{1}\dots i_{n}}(\theta_{1},\dots,\theta_{n})=\left(\prod_{k=1}^{n}m_{i_{k}}L\cosh\theta_{k}\right)\left(1+O(1/L)\right)\]
The leading factor can only be zero when $\theta_{k}=\frac{i\pi}{2}$
for some $k$, which corresponds to $p_{k}=im_{k}$, giving $\mu''=m_{k}$
in case this is the closest singularity. That gives a correction\[
\mathrm{e}^{-m_{k}L}\]
which is the same as the contribution by an on-shell propagator wound
around the finite volume, and such corrections are already included
in the $\mathrm{e}^{-\mu L}$ term of (\ref{eq:corrfinvoldiff}).
Another possibility is that some phase-shift function $\delta(\theta)$
in the $O(1/L)$ terms contributes a large term, which balances the
$1/L$ prefactor. For that its argument must be close to a singularity,
and then according to eqn. (\ref{eq:Smatpole}) we can write\[
\delta(\theta)\sim\log(\theta-u)\sim O(L)\]
where $u$ is the position of the singularity in the phase-shift%
\footnote{Strictly speaking, this argument only works in an integrable field
theory where we know the form (\ref{eq:byjacobian}) of the determinant,
and also know the singularity structure of the phase-shift $\delta$:
the two-particle $S$ matrix amplitude as a function of the rapidity
can only have poles (possibly of higher order, but that does not affect
the main line of reasoning here).%
}. This requires that the singularity is approached exponentially close
(as a function of the volume $L$), but the positions of all these
singularities are again determined by singularities of the vertex
functions, so this gives no new possibilities for the exponent $\mu''$. 

A further issue that can be easily checked is whether the Fourier
integral is convergent for large momenta; the function (\ref{eq:hfunction})
is cut off at the infinities by the factor\[
\prod_{k=1}^{n}\exp(-m_{i_{k}}r\cosh\theta_{k})\]
which can only go wrong if for some $k$\[
\mathrm{\Re e}\,\cosh\theta_{k}<0\]
but that requires \[
\mathrm{\Im m}\,\theta_{k}>\frac{\pi}{2}\]
which is already farther from the real momentum domain then the position
of the on-shell propagator singularity. 

Obtaining a sound mathematical proof requires applying Lüscher's finite
volume expansion directly to form factors which is out of the scope
of the present work. The corrections computed in subsection 4.1 can
be considered as the simplest examples, but we intend to take this
line of investigation further in the future \cite{pozsgayprep}.

We also remark that there are no finite volume states for which the
quantum numbers of any two of the particles are identical. The reason
is that \[
S_{ii}(0)=-1\]
(with the exception of free bosonic theories) and so the wave function
corresponding to the appropriate solution of the Bethe-Yang equations
(\ref{eq:betheyang}) vanishes. We can express this in terms of form
factors as follows:\[
\langle0\vert\mathcal{O}(0,0)\vert\{ I_{1},I_{2},\dots,I_{n}\}\rangle_{i_{1}\dots i_{n},L}=0\]
whenever $I_{k}=I_{l}$ and $i_{k}=i_{l}$ for some $k$ and $l$.
Using this convention we can assume that the summation in (\ref{eq:finitevolspectralrepr})
runs over all possible values of the quantum numbers without exclusions.
Note that even in this case the relation (\ref{eq:ffrelation}) can
be maintained since due to the exchange axiom (\ref{eq:exchangeaxiom})\[
F_{n}^{\mathcal{O}}(\tilde{\theta}_{1},\dots,\tilde{\theta}_{n})_{i_{1}\dots i_{n}}=0\]
whenever $\tilde{\theta}_{k}=\tilde{\theta}_{l}$ and $i_{k}=i_{l}$
for some $k$ and $l$.

It is also worthwhile to mention that there is no preferred way to
order the rapidities on the circle, since there are no genuine asymptotic
\emph{in/out} particle configurations. This means that in relation
(\ref{eq:ffrelation}) there is no preferred way to order the rapidities
inside the infinite volume form factor function $F_{n}^{\mathcal{O}}$.
Different orderings are related by $S$-matrix factors according to
the exchange axiom (\ref{eq:exchangeaxiom}), which are indeed phases.
Such phases do not contribute to correlation functions (cf. the spectral
representation (\ref{eq:spectralrepr})), nor to any physically meaningful
quantity derived from them. In subsection \ref{sub:sly2pff} we show
that relations like (\ref{eq:ffrelation}) must always be understood
to hold only up to physically irrelevant phase factors.

The quantity $\rho_{i_{1}\dots i_{n}}(\theta_{1},\dots,\theta_{n})$
is nothing else than the density of states in rapidity space. It is
also worthwhile to mention that relation (\ref{eq:ffrelation}) can
be interpreted as an expression for the finite volume multi-particle
state in terms of the corresponding infinite volume state as follows\begin{equation}
\vert\{ I_{1},\dots,I_{n}\}\rangle_{i_{1}\dots i_{n},L}=\frac{1}{\sqrt{\rho_{i_{1}\dots i_{n}}(\tilde{\theta}_{1},\dots,\tilde{\theta}_{n})}}\vert\tilde{\theta}_{1},\dots,\tilde{\theta}_{n}\rangle_{i_{1}\dots i_{n}}\label{eq:staterenorm}\end{equation}
This relation between the density and the normalization of states
is a straightforward application of the ideas put forward by Saleur
in \cite{saleurfiniteT}. Using the crossing formula (\ref{eq:ffcrossing}),
eqn. (\ref{eq:staterenorm}) allows us to construct the general form
factor functions (\ref{eq:genff}) in finite volume as follows:\begin{eqnarray}
 &  & \,_{j_{1}\dots j_{m}}\langle\{ I_{1}',\dots,I_{m}'\}\vert\mathcal{O}(0,0)\vert\{ I_{1},\dots,I_{n}\}\rangle_{i_{1}\dots i_{n},L}=\nonumber \\
 &  & \qquad\frac{F_{m+n}^{\mathcal{O}}(\tilde{\theta}_{m}'+i\pi,\dots,\tilde{\theta}_{1}'+i\pi,\tilde{\theta}_{1},\dots,\tilde{\theta}_{n})_{j_{m}\dots j_{1}i_{1}\dots i_{n}}}{\sqrt{\rho_{i_{1}\dots i_{n}}(\tilde{\theta}_{1},\dots,\tilde{\theta}_{n})\rho_{j_{1}\dots j_{m}}(\tilde{\theta}_{1}',\dots,\tilde{\theta}_{m}')}}+O(\mathrm{e}^{-\mu L})\label{eq:genffrelation}\end{eqnarray}
provided that there are no rapidities that are common between the
left and the right states i.e. the sets $\left\{ \tilde{\theta}_{1},\dots,\tilde{\theta}_{n}\right\} $
and $\left\{ \tilde{\theta}_{1}',\dots,\tilde{\theta}_{m}'\right\} $
are disjoint. The latter condition is necessary to eliminate disconnected
pieces.

We stress that eqns. (\ref{eq:ffrelation}, \ref{eq:genffrelation})
are exact to all orders of powers in $1/L$; we refer to the corrections
non-analytic in $1/L$ (eventually decaying exponentially as indicated)
as \emph{residual finite size effects}, following the terminology
introduced in \cite{takacspozsgay}.

\section{Form factors from truncated conformal space}

\subsection{Scaling Lee-Yang model}

\subsubsection{Truncated conformal space approach for scaling Lee-Yang model}

We use the truncated conformal space approach (TCSA) developed by
Yurov and Zamolodchikov in \cite{yurov_zamolodchikov}. The ultraviolet
conformal field theory has central charge $c=-22/5$ and a unique
nontrivial primary field $\Phi$ with scaling weights $\Delta=\bar{\Delta}=-1/5$.
The cylinder of circumference $L$ can be mapped unto the complex
plane using\begin{equation}
z=\exp\frac{2\pi}{L}(\tau-ix)\qquad,\qquad\bar{z}=\exp\frac{2\pi}{L}(\tau+ix)\label{eq:exponentialmap}\end{equation}
The field $\Phi$ is normalized so that it has the following operator
product expansion:\begin{equation}
\Phi(z,\bar{z})\Phi(0,0)=\mathcal{C}(z\bar{z})^{1/5}\Phi(0,0)+(z\bar{z})^{2/5}\mathbb{I}+\dots\label{eq:lyconfope}\end{equation}
where $\mathbb{I}$ is the identity operator and the only nontrivial
structure constant is \[
\mathcal{C}=1.911312699\dots\times i\]
The Hilbert space of the conformal model is given by\[
\mathcal{H}_{LY}=\bigoplus_{h=0,-1/5}\mathcal{V}_{h}\otimes\bar{\mathcal{V}}_{h}\]
where $\mathcal{V}_{h}$ ($\bar{\mathcal{V}}_{h}$) denotes the irreducible
representation of the left (right) Virasoro algebra with highest weight
$h$. 

The Hamiltonian of scaling Lee-Yang model takes the following form
in the perturbed conformal field theory framework:\begin{equation}
H^{SLY}=H_{0}^{LY}+i\lambda\int_{0}^{L}dx\Phi(0,x)\label{eq:lypcftham}\end{equation}
where \[
H_{0}^{LY}=\frac{2\pi}{L}\left(L_{0}+\bar{L}_{0}-\frac{c}{12}\right)\]
is the conformal Hamiltonian. When $\lambda>0$ the theory above has
a single particle in its spectrum with mass $m$ that can be related
to the coupling constant as \cite{lytba} \begin{equation}
\lambda=0.09704845636\dots\times m^{12/5}\label{eq:lymassgap}\end{equation}
and the bulk energy density is given by\begin{equation}
\mathcal{B}=-\frac{\sqrt{3}}{12}m^{2}\label{eq:lybulk}\end{equation}
The $S$-matrix reads \cite{CM}\begin{equation}
S_{LY}(\theta)=\frac{\sinh\theta+i\sin\frac{2\pi}{3}}{\sinh\theta-i\sin\frac{2\pi}{3}}\label{eq:Smatly}\end{equation}
and the particle occurs as a bound state of itself at $\theta=2\pi i/3$
with the three-particle coupling given by\[
\Gamma^{2}=-2\sqrt{3}\]
where the negative sign is due to the nonunitarity of the model. In
this model we define the phase-shift via the relation\begin{equation}
S_{LY}(\theta)=-\mathrm{e}^{i\delta(\theta)}\label{eq:lydeltachoice}\end{equation}
so that $\delta(0)=0$. This means a redefinition of Bethe quantum
numbers $I_{k}$ in the Bethe-Yang equations (\ref{eq:byjacobian})
such they become half-integers for states composed of an even number
of particles; it also means that in the large volume limit, particle
momenta become\[
m\sinh\tilde{\theta}_{k}=\frac{2\pi I_{k}}{L}\]
Due to translational invariance of the Hamiltonian (\ref{eq:lypcftham}),
the conformal Hilbert space $\mathcal{H}$ can be split into sectors
characterized by the eigenvalues of the total spatial momentum \[
P=\frac{2\pi}{L}\left(L_{0}-\bar{L}_{0}\right)\]
the operator $L_{0}-\bar{L}_{0}$ generates Lorentz transformations
and its eigenvalue is called Lorentz spin. For a numerical evaluation
of the spectrum, the Hilbert space is truncated by imposing a cut
in the conformal energy. The truncated conformal space corresponding
to a given truncation and fixed value $s$ of the Lorentz spin reads\[
\mathcal{H}_{\mathrm{TCS}}(s,e_{\mathrm{cut}})=\left\{ |\psi\rangle\in\mathcal{H}\:|\;\left(L_{0}-\bar{L}_{0}\right)|\psi\rangle=s|\psi\rangle,\;\left(L_{0}+\bar{L}_{0}-\frac{c}{12}\right)|\psi\rangle=e|\psi\rangle\,:\, e\leq e_{\mathrm{cut}}\right\} \]
On this subspace, the dimensionless Hamiltonian matrix can be written
as\begin{equation}
h_{ij}=\frac{2\pi}{l}\left(L_{0}+\bar{L}_{0}-\frac{c}{12}+i\frac{\kappa l^{2-2\Delta}}{(2\pi)^{1-2\Delta}}G^{(s)-1}B^{(s)}\right)\label{eq:dimlesstcsaham}\end{equation}
where energy is measured in units of the particle mass $m$, $l=mL$
is the dimensionless volume parameter, \begin{equation}
G_{ij}^{(s)}=\langle i|j\rangle\label{eq:Gs}\end{equation}
is the conformal inner product matrix and \begin{equation}
B_{ij}^{(s)}=\left.\langle i|\Phi(z,\bar{z})|j\rangle\right|_{z=\bar{z}=1}\label{eq:Bs}\end{equation}
is the matrix element of the operator $\Phi$ at the point $z=\bar{z}=1$
on the complex plane between vectors $|i\rangle$, $|j\rangle$ from
$\mathcal{H}_{\mathrm{TCS}}(s,e_{\mathrm{cut}})$. The natural basis
provided by the action of Virasoro generators is not orthonormal and
therefore $G^{(s)-1}$ must be inserted to transform the left vectors
to the dual basis. The Hilbert space and the matrix elements are constructed
using an algorithm developed by Kausch et al. and first used in \cite{ktw}.

Diagonalizing the matrix $h_{ij}$ we obtain the energy levels as
functions of the volume, with energy and length measured in units
of $m$. The maximum value of the cutoff $e_{\mathrm{cut}}$ we used
was $30$, in which case the Hilbert space contains around one thousand
vectors, slightly depending on the spin.

\subsubsection{Exact form factors of the primary field $\Phi$}

Form factors of the trace of the stress-energy tensor $\Theta$ were
computed by Al.B. Zamolodchikov in \cite{Z1}, and using the relation\begin{equation}
\Theta=i\lambda\pi(1-\Delta)\Phi\label{eq:thetaphirel}\end{equation}
we can rewrite them in terms of $\Phi$. They have the form\begin{equation}
F_{n}(\theta_{1},\dots,\theta_{n})=\langle\Phi\rangle H_{n}Q_{n}(x_{1},\dots,x_{n})\prod_{i=1}^{n}\prod_{j=i+1}^{n}\frac{f(\theta_{i}-\theta_{j})}{x_{i}+x_{j}}\label{eq:lyff}\end{equation}
with the notations\begin{eqnarray*}
f(\theta) & = & \frac{\cosh\theta-1}{\cosh\theta+1/2}v(i\pi-\theta)v(-i\pi+\theta)\\
v(\theta) & = & \exp\left(2\int_{0}^{\infty}dt\frac{\sinh\frac{\pi t}{2}\sinh\frac{\pi t}{3}\sinh\frac{\pi t}{6}}{t\sinh^{2}\pi t}\mathrm{e}^{i\theta t}\right)\\
x_{i} & = & \mathrm{e}^{\theta_{i}}\qquad,\qquad H_{n}=\left(\frac{3^{1/4}}{2^{1/2}v(0)}\right)^{n}\end{eqnarray*}
The exact vacuum expectation value of the field $\Phi$ is \[
\langle\Phi\rangle=1.239394325\dots\times i\, m^{-2/5}\]
which can be readily obtained using (\ref{eq:lymassgap}, \ref{eq:thetaphirel})
and also the known vacuum expectation value of $\Theta$ \cite{Z1}\[
\langle\Theta\rangle=-\frac{\pi m^{2}}{4\sqrt{3}}\]
The functions $Q_{n}$ are symmetric polynomials in the variables
$x_{i}$. Defining the elementary symmetric polynomials of $n$ variables
by the relations\[
\prod_{i=1}^{n}(x+x_{i})=\sum_{i=0}^{n}x^{n-i}\sigma_{i}^{(n)}(x_{1},\dots,x_{n})\qquad,\qquad\sigma_{i}^{(n)}=0\mbox{ for }i>n\]
they can be constructed as\begin{eqnarray*}
Q_{1} & = & 1\qquad,\qquad Q_{2}=\sigma_{1}^{(2)}\qquad,\qquad Q_{3}=\sigma_{1}^{(3)}\sigma_{2}^{(3)}\\
Q_{n} & = & \sigma_{1}^{(n)}\sigma_{n-1}^{(n)}P_{n}\quad,\qquad n>3\\
P_{n} & = & \det\mathcal{M}^{(n)}\quad\mbox{where}\quad\mathcal{M}_{ij}^{(n)}=\sigma_{3i-2j+1}^{(n)}\quad,\quad i,j=1,\dots,n-3\end{eqnarray*}
Note that the one-particle form factor is independent of the rapidity:\begin{equation}
F_{1}^{\Phi}=1.0376434349\dots\times im^{-2/5}\label{eq:ly1pff}\end{equation}

\subsection{Ising model with magnetic perturbation}

The critical Ising model is the described by the conformal field theory
with $c=1/2$ and has two nontrivial primary fields: the spin operator
$\sigma$ with $\Delta_{\sigma}=\bar{\Delta}_{\sigma}=1/16$ and the
energy density $\epsilon$ with $\Delta_{\epsilon}=\bar{\Delta}_{\epsilon}=1/2$.
The magnetic perturbation\[
H=H_{0}^{I}+h\int_{0}^{L}dx\sigma(0,x)\]
is massive (and its physics does not depend on the sign of the external
magnetic field $h$). The spectrum and the exact $S$ matrix is described
by the famous $E_{8}$ factorized scattering theory \cite{e8}, which
contains eight particles $A_{i},\; i=1,\dots,8$ with mass ratios
given by \begin{eqnarray*}
m_{2} & = & 2m_{1}\cos\frac{\pi}{5}\\
m_{3} & = & 2m_{1}\cos\frac{\pi}{30}\\
m_{4} & = & 2m_{2}\cos\frac{7\pi}{30}\\
m_{5} & = & 2m_{2}\cos\frac{2\pi}{15}\\
m_{6} & = & 2m_{2}\cos\frac{\pi}{30}\\
m_{7} & = & 2m_{4}\cos\frac{\pi}{5}\\
m_{8} & = & 2m_{5}\cos\frac{\pi}{5}\end{eqnarray*}
and the mass gap relation is \cite{phonebook}\[
m_{1}=(4.40490857\dots)|h|^{8/15}\]
or\begin{equation}
h=\kappa_{h}m_{1}^{15/8}\qquad,\qquad\kappa_{h}=0.06203236\dots\label{eq:ising_massgap}\end{equation}
The bulk energy density is given by\begin{equation}
B=-0.06172858982\dots\times m_{1}^{2}\label{eq:isingbulk}\end{equation}
We also quote the scattering phase shift of two $A_{1}$ particles:
\begin{equation}
S_{11}(\theta)=\left\{ \frac{1}{15}\right\} _{\theta}\left\{ \frac{1}{3}\right\} _{\theta}\left\{ \frac{2}{5}\right\} _{\theta}\quad,\quad\left\{ x\right\} _{\theta}=\frac{\sinh\theta+i\sin\pi x}{\sinh\theta-i\sin\pi x}\label{eq:s11_ising}\end{equation}
All other amplitudes $S_{ab}$ are determined by the $S$ matrix bootstrap
\cite{e8}; the only one we need later is that of the $A_{1}-A_{2}$
scattering, which takes the form\[
S_{12}(\theta)=\left\{ \frac{1}{5}\right\} _{\theta}\left\{ \frac{4}{15}\right\} _{\theta}\left\{ \frac{2}{5}\right\} _{\theta}\left\{ \frac{7}{15}\right\} _{\theta}\]
To have an unambiguous definition of the quantum numbers $I_{i}$
entering the Bethe-Yang equations (\ref{eq:betheyang}), it is convenient
to define phase shift functions $\delta_{ab}$ which are continuous
and odd functions of the rapidity difference $\theta$; we achieve
this using the following convention:\[
S_{ab}(\theta)=S_{ab}(0)\mathrm{e}^{i\delta_{ab}(\theta)}\]
where $\delta_{ab}$ is uniquely specified by continuity and the branch
choice\[
\delta_{ab}(0)=0\]
and it is an odd function of $\theta$ due to the following property
of the scattering amplitude:\[
S_{ab}(\theta)S_{ab}(-\theta)=1\]
(which also implies $S_{ab}(0)=\pm1$). The above redefinition of
the phase shift compared to the original one in eqn. (\ref{eq:origphaseshift})
contains as a special case the Lee-Yang definition (\ref{eq:lydeltachoice})
and also entails appropriate redefinition of quantum numbers depending
on the sign of $S_{ab}(0)$.

\subsubsection{Truncated fermionic space approach for the Ising model}

The conformal Ising model can be represented as the theory of a massless
Majorana fermion with the action\[
\mathcal{A}_{Ising}=\frac{1}{2\pi}\int d^{2}z\left(\bar{\psi}\partial\bar{\psi}+\psi\bar{\partial}\psi\right)\]
On the conformal plane the model has two sectors, with the mode expansions\[
\psi(z)=\begin{cases}
\sum_{r\in\mathbb{Z}+\frac{1}{2}}b_{r}z^{-r-1/2} & \mbox{ Neveu-Schwarz (NS) sector}\\
\sum_{r\in\mathbb{Z}}b_{r}z^{-r-1/2} & \mbox{ Ramond (R) sector}\end{cases}\]
and similarly for the antiholomorphic field $\bar{\psi}$. The Hilbert
space is the direct sum of a certain projection of the NS and R sectors,
with the Virasoro content\[
\mathcal{H}_{Ising}=\bigoplus_{h=0,\frac{1}{2},\frac{1}{16}}\mathcal{V}_{h}\otimes\bar{\mathcal{V}}_{h}\]
The spin field $\sigma$ connects the NS and R sectors, and its matrix
elements $B_{ij}^{(s)}$ in the sector with a given conformal spin
$s$ (cf. eqn. (\ref{eq:Bs})) can be most conveniently computed in
the fermionic basis using the work of Yurov and Zamolodchikov \cite{tfcsa},
who called this method the truncated fermionic space approach. The
fermionic basis can easily be chosen orthonormal, and thus in this
case the metrics $G^{(s)}$ on the spin subspaces (cf. eqn. (\ref{eq:Gs}))
are all given by unit matrices of appropriate dimension. Apart from
the choice of basis all the calculation proceeds very similarly to
the case of the Lee-Yang model. Energy and volume is measured in units
of the lowest particle mass $m=m_{1}$ and using relation (\ref{eq:ising_massgap})
one can write the dimensionless Hamiltonian in the form (\ref{eq:dimlesstcsaham}).
The highest cutoff we use is $e_{\mathrm{cut}}=30$, in which case
the Hilbert space contains around three thousand vectors (slightly
depending on the value of the spin chosen).

We remark that the energy density operator can be represented in the
fermionic language as \[
\epsilon=\bar{\psi}\psi\]
which makes the evaluation of its matrix elements in the fermionic
basis extremely simple.

\subsubsection{Form factors of the energy density operator $\epsilon$}

The form factors of the operator $\epsilon$ in the $E_{8}$ model
were first calculated in \cite{delfino_simonetti} and their determination
was carried further in \cite{resonances}. The exact vacuum expectation
value of the field $\epsilon$ is given by \cite{vevs}\[
\langle\epsilon\rangle=\epsilon_{h}|h|^{8/15}\qquad,\qquad\epsilon_{h}=2.00314\dots\]
or in terms of the mass scale $m=m_{1}$\begin{equation}
\langle\epsilon\rangle=0.45475\dots\times m\label{eq:epsilonexactvev}\end{equation}
The form factors are not known for the general $n$-particle case
in a closed form, i.e. no formula similar to that in (\ref{eq:lyff})
exists. They can be evaluated by solving the appropriate polynomial
recursion relations derived from the form-factor axioms. We do not
present explicit formulae here; instead we refer to the above papers.
For practical calculations we used the results computed by Delfino,
Grinza and Mussardo, which can be downloaded from the Web in \texttt{Mathematica}
format \cite{isingff}.

Our interest in the Ising model is motivated by the fact that this
is the simplest model in which form factors of an operator different
from the perturbing one are known, and also its spectrum and bootstrap
structure is rather complex, both of which stands in contrast with
the much simpler case of scaling Lee-Yang model.

\subsection{Evaluating matrix elements of a local operator $\mathcal{O}$ in
TCSA}

\subsubsection{Identification of multi-particle states}

Diagonalizing the TCSA Hamiltonian (\ref{eq:dimlesstcsaham}) yields
a set of eigenvalues and eigenvectors at each value of the volume,
but it is not immediately obvious how to select the same state at
different values of the volume. Therefore in order to calculate form
factors it is necessary to identify the states with the corresponding
many-particle interpretation.

Finding the vacuum state is rather simple since it is the lowest lying
state in the spin-0 sector and its energy is given by \[
E_{0}(L)=BL+\dots\]
where the ellipsis indicate residual finite size effects decaying
exponentially fast with volume $L$ and $B$ is the bulk energy density
which in the models we consider is exactly known (\ref{eq:lybulk},
\ref{eq:isingbulk}). One-particle states can be found using that
their energies can be expressed as\[
E_{i}^{(s)}(L)=BL+\sqrt{\left(\frac{2\pi s}{L}\right)^{2}+m_{i}^{2}}+\dots\]
again up to residual finite size effects where $s$ is the spin of
the sector considered and $i$ is the species label (every sector
contains a single one-particle state for each species). 

Higher multi-particle states can be identified by comparing the measured
eigenvalues to the levels predicted by the Bethe-Yang equations. Fixing
species labels $i_{1},\dots,i_{n}$ and momentum quantum numbers $I_{1},\dots,I_{n}$,
eqns. (\ref{eq:betheyang}) can be solved to give the rapidities $\tilde{\theta}_{1},\dots,\tilde{\theta}_{n}$
of the particles as function of the dimensionless volume parameter
$l=mL$. Then the energy of the multi-particle state in question is\[
E_{i_{1}\dots i_{n}}^{(I_{1}\dots I_{n})}(L)=BL+\sum_{k=1}^{n}m_{i_{k}}\cosh\tilde{\theta}_{k}+\dots\]
which can be compared to the spectrum. 

For each state there exists a range of the volume, called the \emph{scaling
region}, where $L$ is large enough so that the omitted residual finite
size effects can be safely neglected and small enough so that the
truncation errors are also negligible. More precisely, the scaling
region for any quantity depending on the volume can be defined as
the volume range in which the residual finite size corrections and
the truncation errors are of the same order of magnitude; since both
sources of error show a dependence on the state and the particular
quantity considered (as well as on the value of the cutoff), so does
the exact position of the scaling region itself.

\begin{figure}
\begin{centering}\psfrag{el}{$e(l)$}\psfrag{l}{$l$}\includegraphics[scale=0.75]{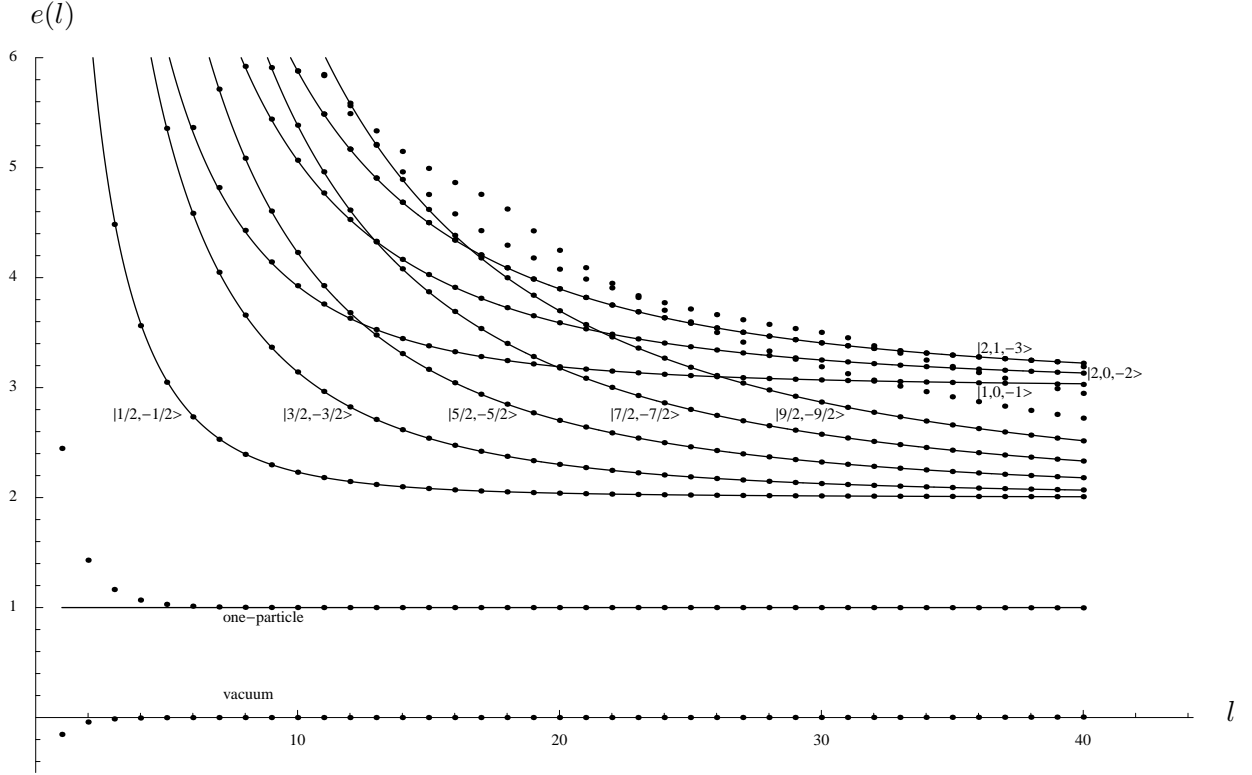}\par\end{centering}

\caption{\label{fig:lyfinvol}The first $13$ states in the finite volume
spectrum of scaling Lee-Yang model. We plot the energy in units of
$m$ (with the bulk subtracted): $e(l)=(E(L)-BL)/m$, against the
dimensionless volume variable $l=mL$. $n$-particle states are labeled
by $|I_{1},\dots,I_{n}\rangle$, where the $I_{k}$ are the momentum
quantum numbers. The state labeled $|2,1,-3\rangle$ is actually two-fold
degenerate because of the presence of $|-2,-1,3\rangle$ (up to a
splitting which vanishes as $\mathrm{e}^{-l}$, cf. the discussion
in subsection 4.3). The dots are the TCSA results and the continuous
lines are the predictions of the Bethe-Yang equations (\ref{eq:betheyang}).
The points not belonging to any of the Bethe-Yang lines drawn are
two- and three-particle states which are only partly contained in
the first $13$ levels due to line crossings, whose presence is a
consequence of the integrability of the model. }
\end{figure}

In the scaling region, we can use a comparison between the Bethe-Yang
predictions and the numerical energy levels to sort the states and
label them by multi-particle quantum numbers. An example is shown
in figure \ref{fig:lyfinvol}, where we plot the first few states
in the spin-0 sector of the scaling Lee-Yang model and their identification
in terms of multi-particle states is given. In this case, the agreement
with the predicted bulk energy density and the Bethe-Yang levels in
the scaling region is better than one part in $10^{4}$ for every
state shown (with the TCSA cutoff taken at $e_{\mathrm{cut}}=30$).

\subsubsection{\label{sub:matrixeval}Evaluation of matrix elements}

Suppose that we computed two Hamiltonian eigenvectors as functions
of the volume $L$ (labeled by their quantum numbers in the Bethe-Yang
description (\ref{eq:betheyang}), omitting the particle species labels
for brevity):\begin{eqnarray*}
|\{ I_{1},\dots,I_{n}\}\rangle_{L} & = & \sum_{i}\Psi_{i}(I_{1},\dots,I_{n};L)|i\rangle\\
|\{ I_{1}',\dots,I_{k}'\}\rangle_{L} & = & \sum_{j}\Psi_{j}(I_{1}',\dots,I_{k}';L)|j\rangle\end{eqnarray*}
in the sector with spin $s$ and spin $s'$, respectively. Let the
inner products of these vectors with themselves be given by\begin{eqnarray*}
\mathcal{N} & = & \sum_{i,j}\Psi_{i}(I_{1},\dots,I_{n};L)G_{ij}^{(s)}\Psi_{j}(I_{1},\dots,I_{n};L)\\
\mathcal{N}' & = & \sum_{i,j}\Psi_{i}(I_{1}',\dots,I_{k}';L)G_{ij}^{(s')}\Psi_{j}(I_{1}',\dots,I_{k}';L)\end{eqnarray*}
It is important that the components of the left eigenvector are not
complex conjugated. In the Ising model we work in a basis where all
matrix and vector components are naturally real. In the Lee-Yang model,
the TCSA eigenvectors are chosen so that all of their components $\Psi_{i}$
are either purely real or purely imaginary depending on whether the
basis vector $|i\rangle$ is an element of the $h=\bar{h}=0$ or the
$h=\bar{h}=-1/5$ component in the Hilbert space. It is well-known
that the Lee-Yang model is non-unitary, which is reflected in the
presence of complex structure constants as indicated in (\ref{eq:lyconfope}).
This particular convention for the structure constants forces upon
us the above inner product, because it is exactly the one under which
TCSA eigenvectors corresponding to different eigenvalues are orthogonal.
We remark that by redefining the structure constants and the conformal
inner product it is also possible to use a manifestly real representation
for the Lee-Yang TCSA (up to some truncation effects that lead to
complex eigenvalues in the vicinity of level crossings \cite{yurov_zamolodchikov}).
Note that the above conventions mean that the phases of the eigenvectors
are fixed up to a sign.

Let us consider a spinless primary field $\mathcal{O}$ with scaling
weights $\Delta_{\mathcal{O}}=\bar{\Delta}_{\mathcal{O}}$, which
can be described as the matrix \[
O_{ij}^{(s',s)}=\left.\langle i|\mathcal{O}(z,\bar{z})|j\rangle\right|_{z=\bar{z}=1}\quad,\quad|i\rangle\in\mathcal{H}_{\mathrm{TCS}}(s',e_{\mathrm{cut}})\quad,\quad|j\rangle\in\mathcal{H}_{\mathrm{TCS}}(s,e_{\mathrm{cut}})\]
between the two truncated conformal space sectors. Then the matrix
element of $\mathcal{O}$ can be computed as\begin{eqnarray}
 &  & m^{-2\Delta_{\mathcal{O}}}\langle\{ I_{1}',\dots,I_{k}'\}\vert\mathcal{O}(0,0)\vert\{ I_{1},\dots,I_{n}\}\rangle_{L}=\nonumber \\
 &  & \qquad\left(\frac{2\pi}{mL}\right)^{2\Delta_{\mathcal{O}}}\frac{1}{\sqrt{\mathcal{N}}}\frac{1}{\sqrt{\mathcal{N}'}}\sum_{j,l}\Psi_{j}(I_{1}',\dots,I_{k}';L)O_{jl}^{(s',s)}\Psi_{l}(I_{1},\dots,I_{n};L)\label{eq:fftcsaevaluation}\end{eqnarray}
where the volume dependent prefactor comes from the transformation
of the primary field $\mathcal{O}$ under the exponential map (\ref{eq:exponentialmap})
and we wrote the equation in a dimensionless form using the mass scale
$m$. The above procedure is a generalization of the one used by Guida
and Magnoli to evaluate vacuum expectation values in \cite{guidamagnoli};
it was extended to one-particle form factors in the context of the
tricritical Ising model by Fioravanti et al. in \cite{onepff}.

\section{Numerical results for elementary form factors}

\subsection{Vacuum expectation values and one-particle form factors}

\subsubsection{\label{sub:opffly} Scaling Lee-Yang model}

Before the one-particle form factor we discuss the vacuum expectation
value. Let us define the dimensionless function\[
\phi(l)=-im^{2/5}\langle0|\Phi|0\rangle_{L}\]
where the finite volume expectation value is evaluated from TCSA using
(\ref{eq:fftcsaevaluation}).%
\begin{figure}
\begin{centering}\psfrag{l}{$l$}
\psfrag{vev}{$\phi$}\includegraphics[scale=1.3]{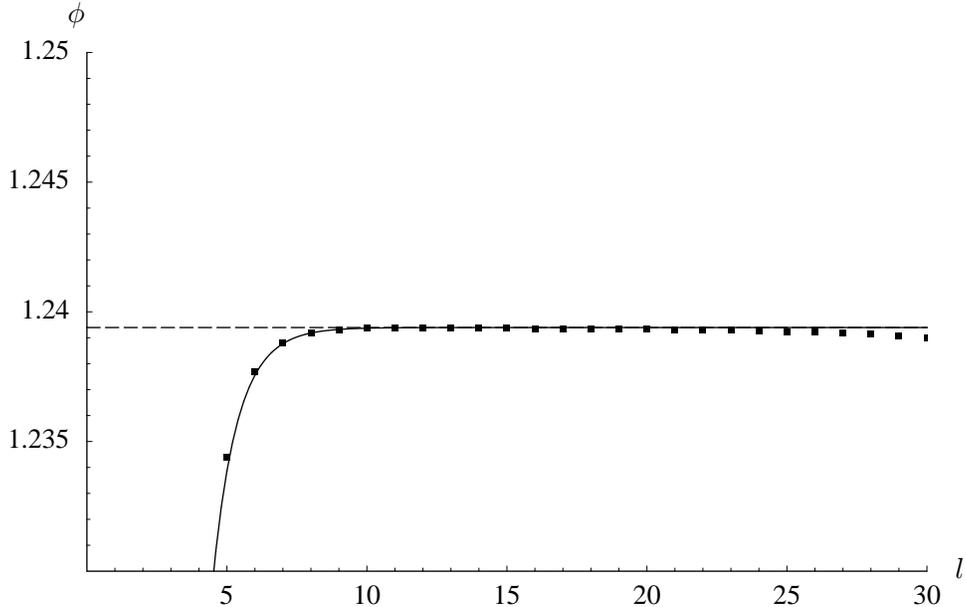}\par\end{centering}

\caption{\label{fig:lyvev}The vacuum expectation value of $\Phi$ in finite
volume. The dashed line shows the exact infinite volume value, while
the continuous line corresponds to eqn. (\ref{eq:vevLcorr}). }
\end{figure}
We performed measurement of $\phi$ as a function of both the cutoff
$e_{\mathrm{cut}}=21\dots30$ and the volume $l=1\dots30$ and then
extrapolated the cutoff dependence fitting a function\[
\phi(l,e_{\mathrm{cut}})=\phi(l)+A(l)e_{\mathrm{cut}}^{-12/5}\]
(where the exponent was chosen by verifying that it provides an optimal
fit to the data). The data corresponding to odd and even values of
the cutoff must be extrapolated separately \cite{takacspozsgay},
therefore one gets two estimates for the result, but they only differ
by a very small amount (of order $10^{-5}$ at $l=30$ and even less
for smaller volumes). The theoretical prediction for $\phi(l)$ is\[
\phi(l)=1.239394325\dots+O(\mathrm{e}^{-l})\]
The numerical result (after extrapolation) is shown in figure \ref{fig:lyvev}
from which it is clear that there is a long scaling region. Estimating
the infinite volume value from the flattest part of the extrapolated
curve (at $l$ around $12$) we obtain the following measured value\[
\phi(l=\infty)=1.23938\dots\]
where the numerical errors from TCSA are estimated to affect only
the last displayed digit, which corresponds to an agreement within
one part in $10^{5}$.

\begin{figure}
\noindent \begin{centering}\psfrag{O}{$\Phi$}
\psfrag{sum}{\Large $+\,\displaystyle\sum_i$}
\psfrag{i}{$i$}\includegraphics{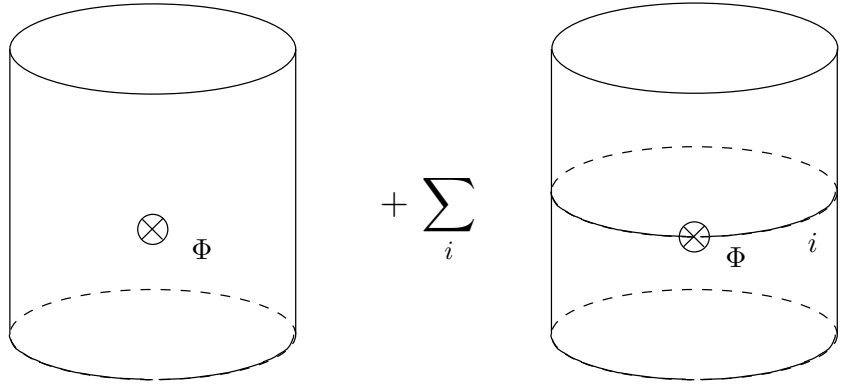}\par\end{centering}

\caption{\label{fig:vevLcorr} Graphical representation of eqn. (\ref{eq:vevLcorr}).}
\end{figure}

There is also a way to compute the leading exponential correction,
which was derived by Delfino \cite{delfino_vev}:\begin{equation}
\langle\Phi\rangle_{L}=\langle\Phi\rangle+\frac{1}{\pi}\sum_{i}F_{2}(i\pi,0)_{ii}K_{0}(m_{i}r)+\dots\label{eq:vevLcorr}\end{equation}
where \[
K_{0}(x)=\int_{0}^{\infty}d\theta\,\cosh\theta\,\mathrm{e}^{-x\cosh\theta}\]
is the modified Bessel-function, and the summation is over the particle
species $i$ (there is only a single term in the scaling Lee-Yang
model). This agrees very well with the numerical data, as demonstrated
in table \ref{tab:vevLcorr} and also in figure \ref{fig:lyvev}.
Using Lüscher's finite-volume perturbation theory introduced in \cite{luscher_onept},
the correction term can be interpreted as the sum of Feynman diagrams
where there is exactly one propagator that winds around the cylinder,
and therefore eqn. (\ref{eq:vevLcorr}) can be represented graphically
as shown in figure \ref{fig:vevLcorr}.

\begin{table}
\noindent \begin{centering}\begin{tabular}{|c|c|c|}
\hline 
$l$&
$\phi(l)$ (predicted)&
$\phi(l)$ (TCSA)\tabularnewline
\hline
\hline 
2&
1.048250&
1.112518\tabularnewline
\hline 
3&
1.184515&
1.195345\tabularnewline
\hline 
4&
1.222334&
1.224545\tabularnewline
\hline 
5&
1.233867&
1.234396\tabularnewline
\hline 
6&
1.237558&
1.237698\tabularnewline
\hline 
7&
1.238774&
1.238811\tabularnewline
\hline 
8&
1.239182&
1.239189\tabularnewline
\hline 
9&
1.239321&
1.239317\tabularnewline
\hline 
10&
1.239369&
1.239360\tabularnewline
\hline 
11&
1.239385&
1.239373\tabularnewline
\hline
12&
1.239391&
1.239375\tabularnewline
\hline
\end{tabular}\par\end{centering}

\caption{\label{tab:vevLcorr} Comparison of eqn. (\ref{eq:vevLcorr}) to
TCSA data}
\end{table}

\begin{figure}
\begin{centering}\psfrag{f1}{$\tilde{f}_1$}\psfrag{l}{$l$}
\psfrag{s=0}{$s=0$}
\psfrag{s=1}{$s=1$}
\psfrag{s=2}{$s=2$}
\psfrag{Exact}{exact}\includegraphics[scale=1.3]{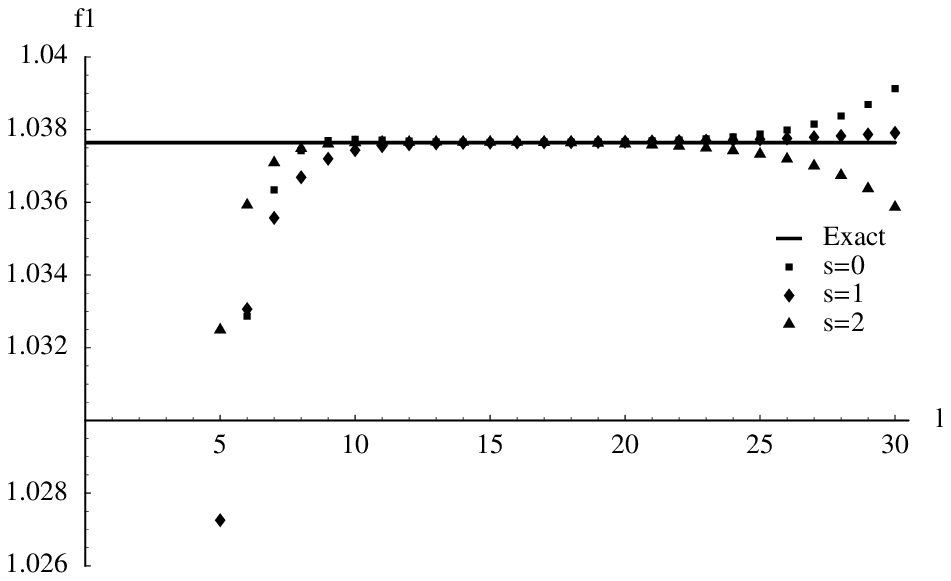}\par\end{centering}

\caption{\label{fig:oply}One-particle form factor from sectors with spin
$s=0,1,2$. The continuous line shows the exact infinite volume prediction.}
\end{figure}

To measure the one-particle form factor we use the correspondence
(\ref{eq:ffrelation}) between the finite and infinite volume form
factors to define the dimensionless function\[
\tilde{f}_{1}^{s}(l)=-im^{2/5}\left(l^{2}+\left(2\pi s\right)^{2}\right)^{1/4}\langle0|\Phi|\{ s\}\rangle_{L}\]
where $|\{ s\}\rangle_{L}$ is the finite volume one-particle state
with quantum number $I=s$ i.e. from the spin-$s$ sector. The theoretical
prediction for this quantity is\begin{equation}
\tilde{f}_{1}^{s}(l)=1.0376434349\dots+O(\mathrm{e}^{-l})\label{eq:lyopffpred}\end{equation}
The numerical results (after extrapolation in the cutoff) are shown
in figure \ref{fig:oply}. The scaling region gives the following
estimates for the infinite volume limit:\begin{eqnarray*}
\tilde{f}_{1}^{0}(l=\infty) & = & 1.037654\dots\\
\tilde{f}_{1}^{1}(l=\infty) & = & 1.037650\dots\\
\tilde{f}_{1}^{2}(l=\infty) & = & 1.037659\dots\end{eqnarray*}
which show good agreement with eqn. (\ref{eq:lyopffpred}) (the relative
deviation is again around $10^{-5}$, as for the vacuum expectation
value).

\subsubsection{Ising model in magnetic field}

\begin{figure}
\noindent \begin{centering}\psfrag{igaziertek}{$\phi(l)$}
\psfrag{l}{$l$}
\psfrag{extrapolated}{extrapolated}
\psfrag{ecut30}{$e_{\mathrm{cut}}=30$}
\psfrag{ecut28}{$e_{\mathrm{cut}}=28$}
\psfrag{ecut26}{$e_{\mathrm{cut}}=26$}\includegraphics[scale=1.2]{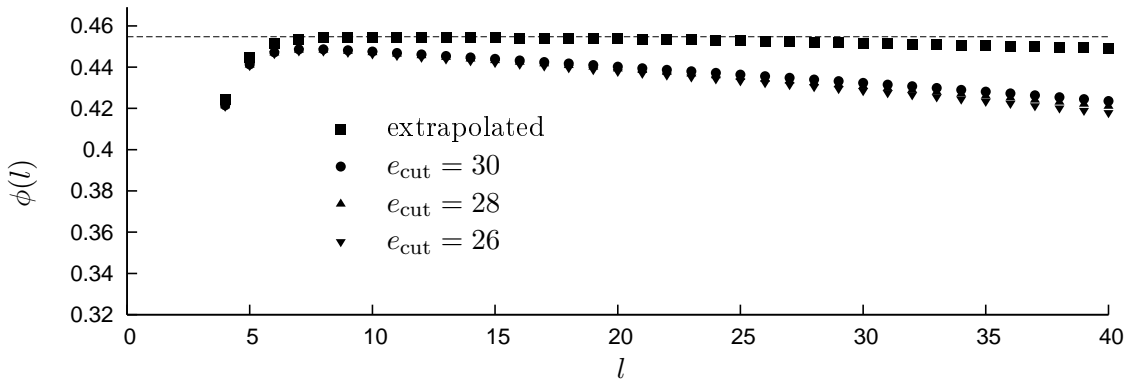}\par\end{centering}

\caption{\label{fig:vevisingraw} Measuring the vacuum expectation value of
$\epsilon$ in the Ising model}
\end{figure}

For the Ising model, we again start with checking the dimensionless
vacuum expectation value for which, using eqn. (\ref{eq:epsilonexactvev})
we have the prediction \[
\phi(l)=\frac{1}{m}\langle\epsilon\rangle_{L}=0.45475\dots+O\left(\mathrm{e}^{-l}\right)\]
where $m=m_{1}$ is the mass of the lightest particle and $l=mL$
as before. The TCSA data are shown in figure \ref{fig:vevisingraw}.
Note that there is substantial dependence on the cutoff $e_{\mathrm{cut}}$
and also that extrapolation in $e_{\mathrm{cut}}$ is really required
to achieve good agreement with the infinite volume limit. Reading
off the plateau value from the extrapolated data gives the estimate\[
\frac{1}{m}\langle\epsilon\rangle=0.4544\dots\]
for the infinite volume vacuum expectation value, which has $8\cdot10^{-4}$
relative deviation from the exact result. Our first numerical comparison
thus already tells us that we can expect much larger truncation errors
than in the Lee-Yang case. It is also clear from figure \ref{fig:vevisingraw}
that in order to attain suitable precision in the Ising model extrapolation
in the cutoff is very important.

\begin{figure}
\noindent \begin{centering}\psfrag{f0}{$\bar{\phi}(l)$}
\psfrag{l}{$l$}
\psfrag{ecut-fitted}{extrapolated}
\psfrag{predicted}{predicted}
\psfrag{infinite-volume}{infinite volume}\includegraphics[scale=1.3]{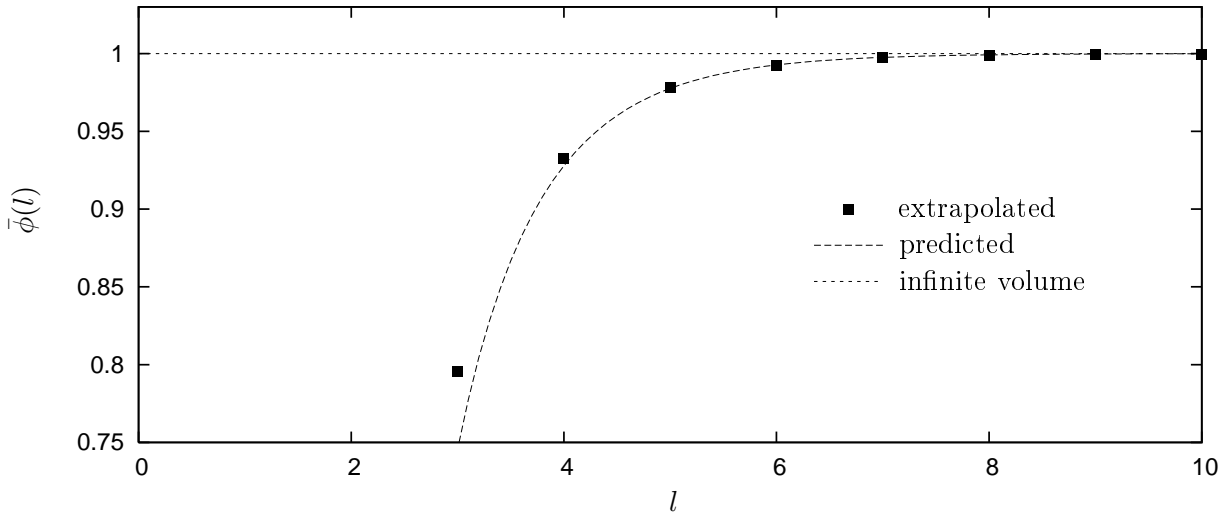}\par\end{centering}

\caption{\label{fig:vevisingLdep}The volume dependence of the vacuum expectation
value of $\epsilon$ in the Ising model, showing the extrapolated
value and the prediction from eqn. (\ref{eq:vevLcorr}), normalized
by the value in the infinite volume limit.}
\end{figure}

Defining the function \[
\bar{\phi}(l)=\langle\epsilon\rangle_{L}/\langle\epsilon\rangle\]
we can calculate the leading exponential correction using eqn. (\ref{eq:vevLcorr})
and the exact two-particle form factors from \cite{isingff}. It only
makes sense to include particles $i=1,2,3$ since the contribution
of the fourth particle is subleading with respect to two-particle
terms from the lightest particle due to $m_{4}>2m_{1}$. The result
is shown in figure \ref{fig:vevisingLdep}; we do not give the data
in numerical tables, but we mention that the relative deviation between
the predicted and measured value is better than $10^{-3}$ in the
range $5<l<10$. 

\begin{figure}
\noindent \begin{centering}\psfrag{f1}{${\tilde f}_1$}
\psfrag{f2}{${\tilde f}_2$}
\psfrag{f3}{${\tilde f}_3$}
\psfrag{l}{$l$}
\psfrag{s=0}{$\scriptstyle s=0$}
\psfrag{s=1}{$\scriptstyle s=1$}
\psfrag{s=2}{$\scriptstyle s=2$}
\psfrag{s=3}{$\scriptstyle s=3$}
\psfrag{exact}{\small exact}\\
\subfigure[$A_1$]{\includegraphics{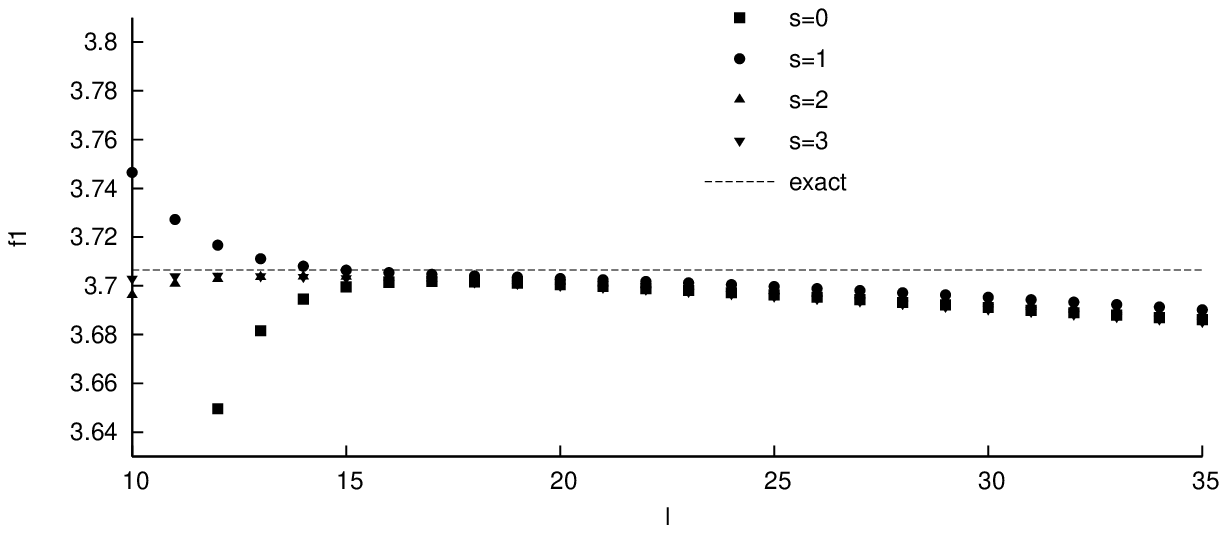}}\\
\subfigure[$A_2$]{\includegraphics{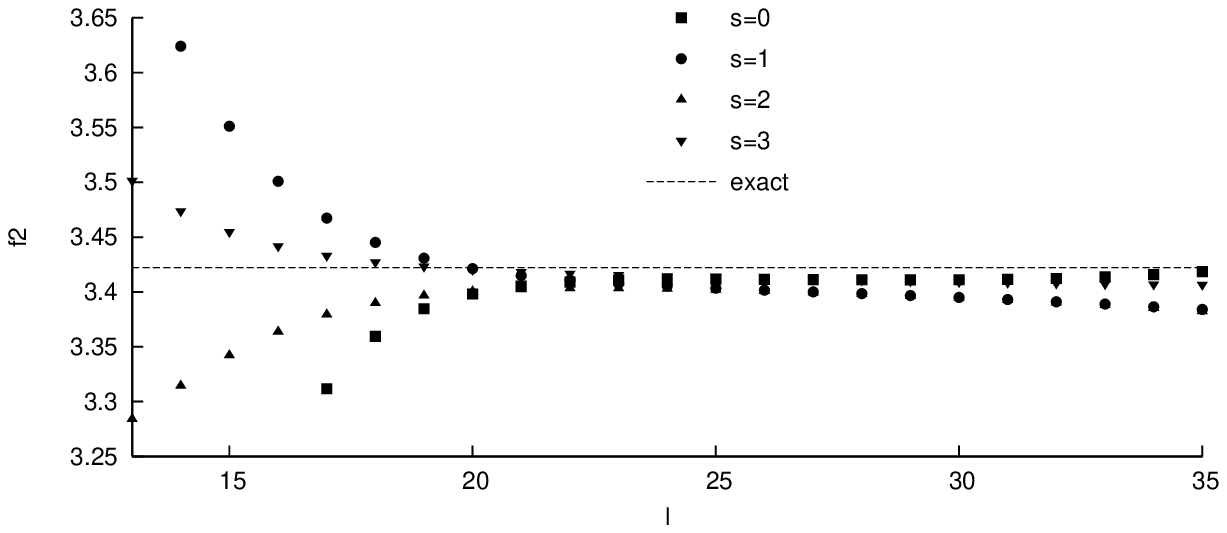}}\\
\subfigure[$A_3$]{\includegraphics{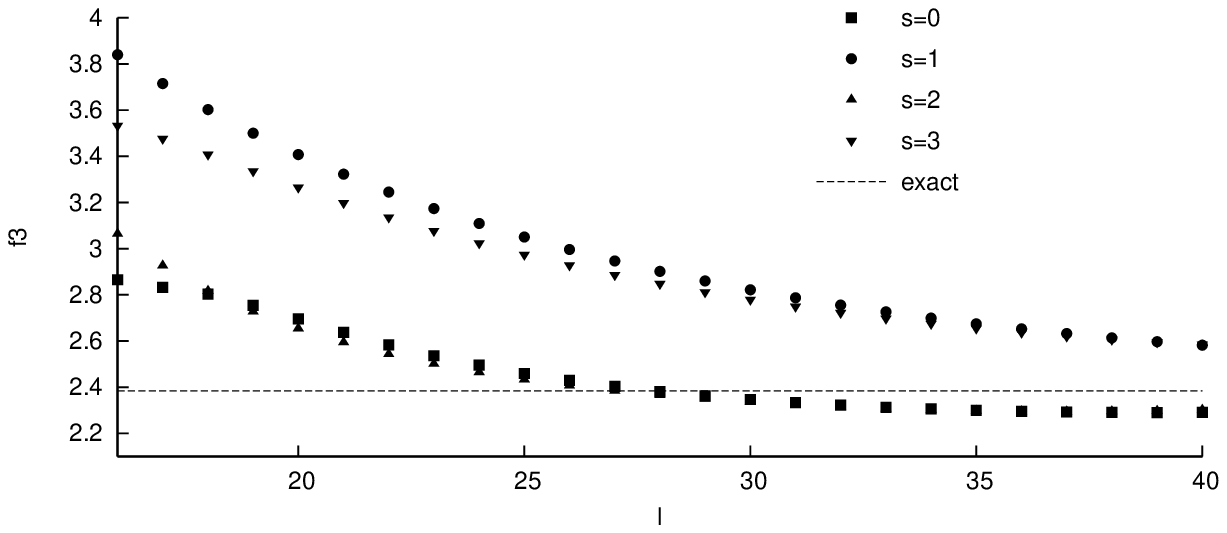}}\par\end{centering}

\caption{\label{fig:onepffising}One-particle form factors measured from TCSA
(dots) compared to the infinite-volume prediction from exact form
factors. All numerical data have been extrapolated to $e_{\mathrm{cut}}=\infty$
and $s$ denotes the Lorentz spin of the state considered. The relative
deviation in the scaling region is around $10^{-3}$ for $A_{1}$
and $A_{2}$; there is no scaling region for $A_{3}$ (see the discussion
in the main text).}
\end{figure}

From now on we normalize all form factors of the operator $\epsilon$
by the infinite volume vacuum expectation value (\ref{eq:epsilonexactvev}),
i.e. we consider form factors of the operator\begin{equation}
\Psi=\epsilon/\langle\epsilon\rangle\label{eq:psidef}\end{equation}
which conforms with the conventions used in \cite{resonances,isingff}.
We define the dimensionless one-particle form factor functions as\[
\tilde{f}_{i}^{s}(l)=\left(\left(\frac{m_{i}l}{m_{1}}\right)^{2}+\left(2\pi s\right)^{2}\right)^{1/4}\langle0|\Psi|\{ s\}\rangle_{i,\, L}\]
In the plots of figure \ref{fig:onepffising} we show how these functions
measured from TCSA compare to predictions from the exact form factors
for particles $i=1,2,3$ and spins $s=0,1,2,3$. 

It is evident that the scaling region sets in much later than for
the Lee-Yang model; therefore for the Ising model we do not plot data
for low values of the volume (all plots start from $l\sim10...15$).
This also means that truncation errors in the scaling region are also
much larger than in the scaling Lee-Yang model; we generally found
errors larger by an order of magnitude after extrapolation in the
cutoff. We remark that extrapolation improves the precision by an
order of magnitude compared to the raw data at the highest value of
the cutoff. 

\begin{figure}
\noindent \begin{centering}\includegraphics{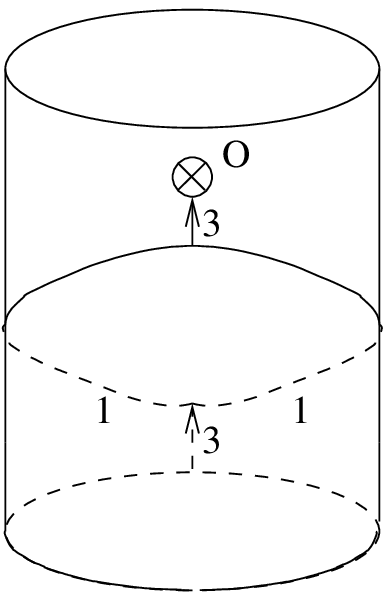}\par\end{centering}

\caption{\label{fig:ising3pcorr}Leading finite size correction (a so-called
$\mu$-term) to the one-particle form factor of $A_{3}$, which results
from the process of splitting up into two copies of $A_{1}$ which
then wind around the cylinder once before recombining into $A_{3}$
again.}
\end{figure}

Note the rather large finite size correction in the case of $A_{3}$.
This can be explained rather simply as the presence of a so-called
$\mu$-term. We can again apply Lüscher's finite-volume perturbation
theory, which we use in the form given by Klassen and Melzer in \cite{klassenmelzer}
for finite volume mass corrections. The generalization to one-particle
matrix elements is straightforward, and for a static particle it gives
the diagram depicted in figure \ref{fig:ising3pcorr}, whose contribution
has the volume dependence\[
\mathrm{e}^{-\mu_{311}L}\qquad,\qquad\mu_{311}=\sqrt{m_{1}^{2}-\frac{m_{3}^{2}}{4}}=0.10453\dots\times m_{1}\]
i.e. we can expect a contribution suppressed only by $\mathrm{e}^{-0.1l}$.
A numerical fit of the $l$-dependence in the $s=0$ case is perfectly
consistent with this expectation. As a result, no scaling region can
be found, because truncation errors are too large in the volume range
where the exponential correction is suitably small. We do not elaborate
on this issue further here; we only mention that starting from this
point there are other interesting observations that can be made, and
we plan to return to them in a separate publication \cite{pozsgayprep}.

\subsection{Two-particle form factors}

\subsubsection{\label{sub:sly2pff} Scaling Lee-Yang model}

Following the ideas in the previous subsection, we can again define
a dimensionless function for each two-particle state as follows:\[
f_{2}(l)_{I_{1}I_{2}}=-im^{2/5}\langle0|\Phi|\{ I_{1},I_{2}\}\rangle_{L}\quad,\quad l=mL\]
Relation (\ref{eq:ffrelation}) gives the following prediction in
terms of the exact form-factors:\begin{equation}
f_{2}(l)_{I_{1}I_{2}}=\frac{-im^{2/5}}{\sqrt{\rho_{11}(\tilde{\theta}_{1}(l),\tilde{\theta}_{2}(l))}}F_{2}^{\Phi}(\tilde{\theta}_{1}(l),\tilde{\theta}_{2}(l))+O(\mathrm{e}^{-l})\label{eq:lyf2prediction}\end{equation}
where $\tilde{\theta}_{1}(l),\,\tilde{\theta}_{2}(l)$ solve the Bethe-Yang
equations\begin{eqnarray*}
l\sinh\tilde{\theta}_{1}+\delta(\tilde{\theta}_{1}-\tilde{\theta}_{2}) & = & 2\pi I_{1}\\
l\sinh\tilde{\theta}_{2}+\delta(\tilde{\theta}_{2}-\tilde{\theta}_{1}) & = & 2\pi I_{2}\end{eqnarray*}
and the density of states is given by\[
\rho_{11}(\theta_{1},\theta_{2})=l^{2}\cosh\theta_{1}\cosh\theta_{2}+l\cosh\theta_{1}\varphi(\theta_{2}-\theta_{1})+l\cosh\theta_{2}\varphi(\theta_{1}-\theta_{2})\]
the phase shift $\delta$ is defined according to eqn. (\ref{eq:lydeltachoice})
and \[
\varphi(\theta)=\frac{d\delta(\theta)}{d\theta}\]
There is a further issue to take into account: the relative phases
of the multi-particle states are a matter of convention and the choice
made in subsection \ref{sub:matrixeval} for the TCSA eigenvectors
may differ from the convention adapted in the form factor bootstrap.
Therefore in the numerical work we compare the absolute values of
the functions $f_{2}(l)$ computed from TCSA with those predicted
from the exact form factors. Note that this issue is present for any
non-diagonal matrix element, and was in fact tacitly dealt with in
the case of one-particle matrix elements treated in subsection \ref{sub:opffly}.

\begin{figure}
\begin{centering}\psfrag{f2}{$|f_2(l)|$}\psfrag{l}{$l$}
\psfrag{tcsa11}{$\scriptstyle\langle 0|\Phi|\{ 1/2,-1/2\}\rangle$}
\psfrag{tcsa33}{$\scriptstyle\langle 0|\Phi|\{ 3/2,-3/2\}\rangle$}
\psfrag{tcsa55}{$\scriptstyle\langle 0|\Phi|\{ 5/2,-5/2\}\rangle$}\includegraphics[scale=1.3]{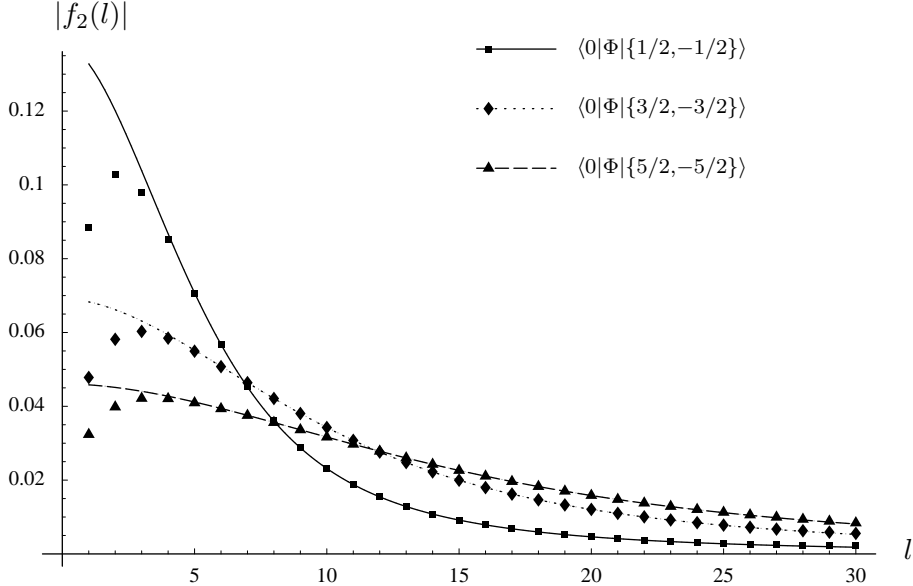}\par\end{centering}

\caption{\label{fig:lytps0}Two-particle form factors in the spin-$0$ sector.
Dots correspond to TCSA data, while the lines show the corresponding
form factor prediction.}
\end{figure}

\begin{figure}
\begin{centering}\psfrag{f2}{$|f_2(l)|$}\psfrag{l}{$l$}
\psfrag{tcsa13}{$\scriptstyle\langle 0|\Phi|\{ 3/2,-1/2\}\rangle$}
\psfrag{tcsa35}{$\scriptstyle\langle 0|\Phi|\{ 5/2,-3/2\}\rangle$}
\psfrag{tcsa15}{$\scriptstyle\langle 0|\Phi|\{ 5/2,-1/2\}\rangle$}
\psfrag{tcsa37}{$\scriptstyle\langle 0|\Phi|\{ 7/2,-3/2\}\rangle$}\includegraphics[scale=1.3]{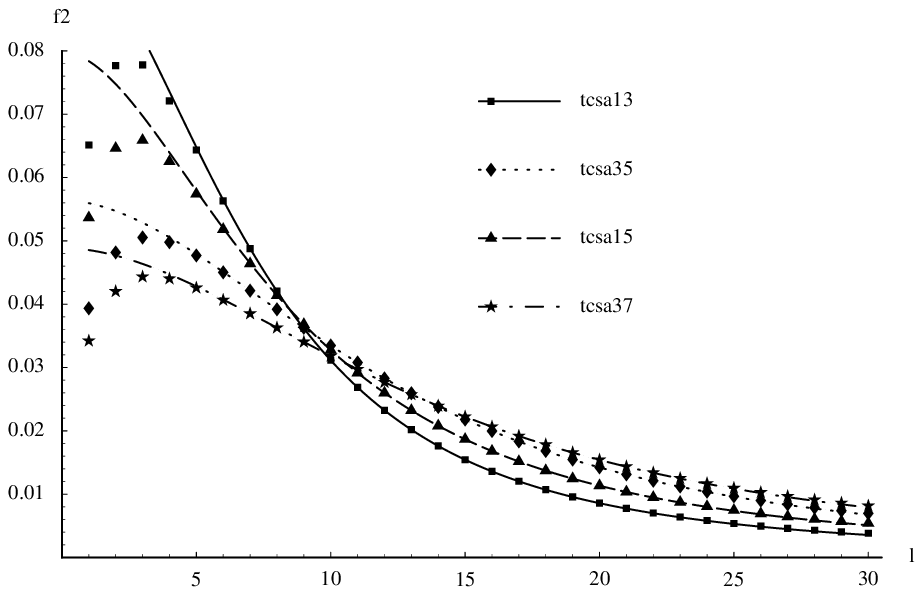}\par\end{centering}

\caption{\label{fig:lytps12}Two-particle form factors in the spin-$1$ and
spin-$2$ sectors. Dots correspond to TCSA data, while the lines show
the corresponding form factor prediction.}
\end{figure}

\begin{figure}
\begin{centering}\psfrag{logdevf2}{$\log |f_2^{FF}(l)-f_2^{TCSA}(l)|$}
\psfrag{l}{$l$}\includegraphics{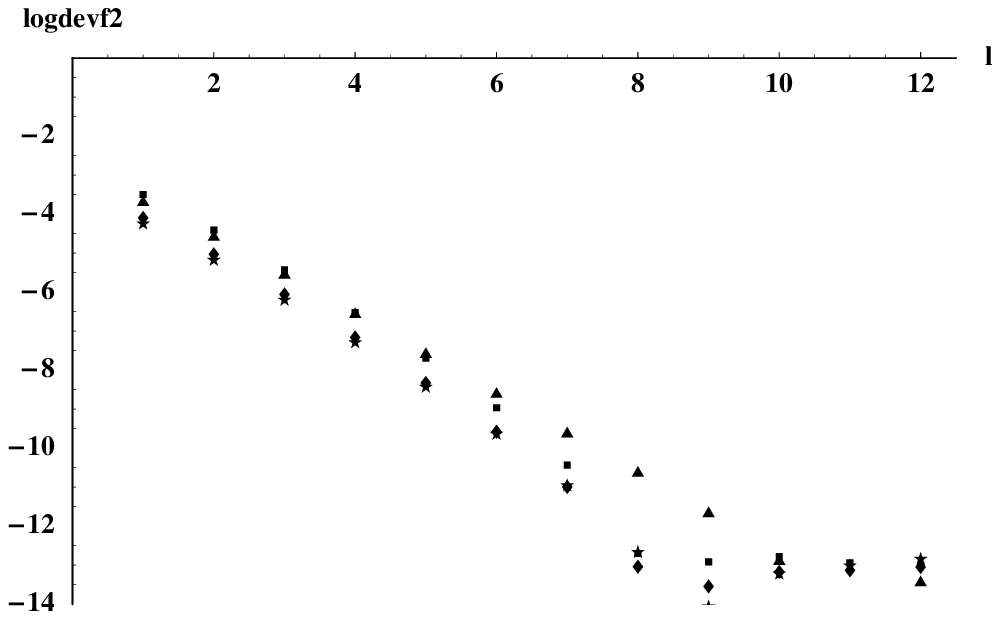}\par\end{centering}

\caption{\label{fig:errorterm}Estimating the error term in (\ref{eq:lyf2prediction})
using the data in table \ref{tab:lytps12}. The various plot symbols
correspond to the same states as specified in figure \ref{fig:lytps12}.}
\end{figure}

The prediction (\ref{eq:lyf2prediction}) for the finite volume two-particle
form factors is compared with spin-$0$ states graphically in figure
\ref{fig:lytps0} and numerically in table \ref{tab:lytps0}, while
the spin-$1$ and spin-$2$ case is presented in figure \ref{fig:lytps12}
and in table \ref{tab:lytps12}. These contain no more than a representative
sample of our data: we evaluated similar matrix elements for a large
number of two-particle states for values of the volume parameter $l$
running from $1$ to $30$. The behaviour of the relative deviation
is consistent with the presence of a correction of $e^{-l}$ type
up to $l\sim9\dots10$ (i.e. the logarithm of the deviation is very
close to being a linear function of $l$), and after $l\sim16\dots18$
it starts to increase due to truncation errors. This is demonstrated
in figure \ref{fig:errorterm} using the data presented in table \ref{tab:lytps12}
for spin-$1$ and spin-$2$ states%
\footnote{Note that the dependence of the logarithm of the deviation on the
volume is not exactly linear because the residual finite size correction
can also contain a factor of some power of $l$, and so it is expected
that a $\log l$ contribution is also present in the data plotted
in figure \ref{fig:errorterm}.%
}, but it is equally valid for all the other states we examined. In
the intermediate region $l\sim10\dots16$ the two sources of numerical
deviation are of the same order, and so that range can be considered
as the optimal scaling region: according to the data in the tables
agreement there is typically around $10^{-4}$ (relative deviation).
It is also apparent that scaling behaviour starts at quite low values
of the volume (around $l\sim4$ the relative deviation is already
down to around 1\%). 

It can be verified by explicit evaluation that in the scaling region
the Bethe-Yang density of states ($\rho$) given in (\ref{eq:byjacobian})
differs by corrections of relative magnitude $10^{-1}-10^{-2}$ (analytically:
of order $1/l$) from the free density of states ($\rho^{0}$) in
(\ref{eq:freejacobidet}), and therefore without using the proper
interacting density of states it is impossible to obtain the precision
agreement we demonstrated. In fact the observed $10^{-4}$ relative
deviation corresponds to corrections of order $l^{-4}$ at $l=10$,
but it is of the order of estimated truncation errors%
\footnote{Truncation errors can be estimated by examining the dependence of
the extracted data on the cutoff $e_{\mathrm{cut}}$, as well as by
comparing TCSA energy levels to the Bethe-Yang predictions.%
}. 

These results are very strong evidence for the main statement in (\ref{eq:lyf2prediction})
(and thus also (\ref{eq:ffrelation})), namely, that all $1/L$ corrections
are accounted by the proper interacting state density factor and that
all further finite size corrections are just residual finite size
effects decaying exponentially in $L$. In section \ref{sub:manypff}
we show that data from higher multi-particle form factors fully support
the above conclusions drawn from the two-particle form factors.

\begin{table}
\begin{centering}\begin{tabular}{|r||c|c||c|c||c|c|}
\hline 
&
\multicolumn{2}{c|}{$I_{1}=1/2$ , $I_{2}=-1/2$}&
\multicolumn{2}{c|}{$I_{1}=3/2$ , $I_{2}=-3/2$}&
\multicolumn{2}{c|}{$I_{1}=5/2$ , $I_{2}=-5/2$}\tabularnewline
\hline 
$l$&
TCSA&
FF&
TCSA&
FF&
TCSA&
FF\tabularnewline
\hline
\hline 
2&
0.102780&
0.120117&
0.058158&
0.066173&
0.039816&
0.045118\tabularnewline
\hline 
4&
0.085174&
0.086763&
0.058468&
0.059355&
0.042072&
0.042729\tabularnewline
\hline 
6&
0.056828&
0.056769&
0.050750&
0.050805&
0.039349&
0.039419\tabularnewline
\hline 
8&
0.036058&
0.035985&
0.042123&
0.042117&
0.035608&
0.035614\tabularnewline
\hline 
10&
0.023168&
0.023146&
0.034252&
0.034248&
0.031665&
0.031664\tabularnewline
\hline 
12&
0.015468&
0.015463&
0.027606&
0.027604&
0.027830&
0.027828\tabularnewline
\hline
14&
0.010801&
0.010800&
0.022228&
0.022225&
0.024271&
0.024267\tabularnewline
\hline
16&
0.007869&
0.007867&
0.017976&
0.017972&
0.021074&
0.021068\tabularnewline
\hline
18&
0.005950&
0.005945&
0.014652&
0.014645&
0.018268&
0.018258\tabularnewline
\hline
20&
0.004643&
0.004634&
0.012061&
0.012050&
0.015844&
0.015827\tabularnewline
\hline
\end{tabular}\par\end{centering}

\caption{\label{tab:lytps0}Two-particle form factors $\left|f_{2}(l)\right|$
in the spin-$0$ sector}
\end{table}

\begin{table}
\begin{centering}\begin{tabular}{|r||c|c||c|c||c|c||c|c|}
\hline 
&
\multicolumn{2}{c|}{$I_{1}=3/2$ , $I_{2}=-1/2$}&
\multicolumn{2}{c|}{$I_{1}=5/2$ , $I_{2}=-3/2$}&
\multicolumn{2}{c|}{$I_{1}=5/2$ , $I_{2}=-1/2$}&
\multicolumn{2}{c|}{$I_{1}=7/2$ , $I_{2}=-3/2$}\tabularnewline
\hline 
$l$&
TCSA&
FF&
TCSA&
FF&
TCSA&
FF&
TCSA&
FF\tabularnewline
\hline
\hline 
2&
0.077674&
0.089849&
0.048170&
0.054711&
0.064623&
0.074763&
0.042031&
0.047672\tabularnewline
\hline 
4&
0.072104&
0.073571&
0.049790&
0.050566&
0.062533&
0.063932&
0.044034&
0.044716\tabularnewline
\hline 
6&
0.056316&
0.056444&
0.045031&
0.045100&
0.051828&
0.052009&
0.040659&
0.040724\tabularnewline
\hline 
8&
0.042051&
0.042054&
0.039191&
0.039193&
0.041370&
0.041394&
0.036284&
0.036287\tabularnewline
\hline 
10&
0.031146&
0.031144&
0.033469&
0.033467&
0.032757&
0.032759&
0.031850&
0.031849\tabularnewline
\hline 
12&
0.023247&
0.023245&
0.028281&
0.028279&
0.026005&
0.026004&
0.027687&
0.027684\tabularnewline
\hline
14&
0.017619&
0.017616&
0.023780&
0.023777&
0.020802&
0.020799&
0.023941&
0.023936\tabularnewline
\hline
16&
0.013604&
0.013599&
0.019982&
0.019977&
0.016808&
0.016802&
0.020659&
0.020652\tabularnewline
\hline
18&
0.010717&
0.010702&
0.016831&
0.016822&
0.013735&
0.013724&
0.017835&
0.017824\tabularnewline
\hline
20&
0.008658&
0.008580&
0.014249&
0.014227&
0.011357&
0.011337&
0.015432&
0.015413\tabularnewline
\hline
\end{tabular}\par\end{centering}

\caption{\label{tab:lytps12}Two-particle form factors $\left|f_{2}(l)\right|$
in the spin-$1$ and spin-$2$ sectors}
\end{table}

\subsubsection{Ising model in magnetic field}

\begin{figure}
\noindent \begin{centering}\psfrag{f11}{$|\bar{f}_{11}|$}\psfrag{l}{$l$}
\psfrag{---a-0.5a0.5}{$\scriptstyle\langle 0|\Psi|\{1/2,-1/2\}\rangle_{11}$}
\psfrag{---a-1.5a1.5}{$\scriptstyle\langle 0|\Psi|\{3/2,-3/2\}\rangle_{11}$}
\psfrag{---a-0.5a1.5}{$\scriptstyle\langle 0|\Psi|\{3/2,-1/2\}\rangle_{11}$}\subfigure[$A_1A_1$]{\includegraphics[scale=1.2]{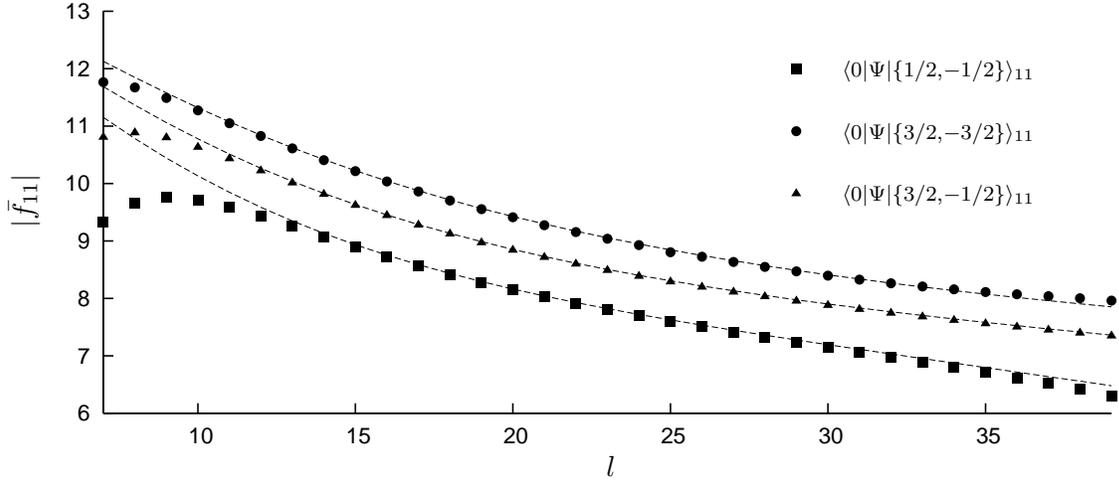}}\\
\psfrag{f12}{$|\bar{f}_{12}|$}\psfrag{l}{$l$}
\psfrag{---a-1b1}{$\scriptstyle\langle 0|\Psi|\{-1,1\}\rangle_{12}$}
\psfrag{---a0b1}{$\scriptstyle\langle 0|\Psi|\{0,1\}\rangle_{12}$}
\psfrag{---a1b1}{$\scriptstyle\langle 0|\Psi|\{1,1\}\rangle_{12}$}\subfigure[$A_1A_2$]{\includegraphics[scale=1.2]{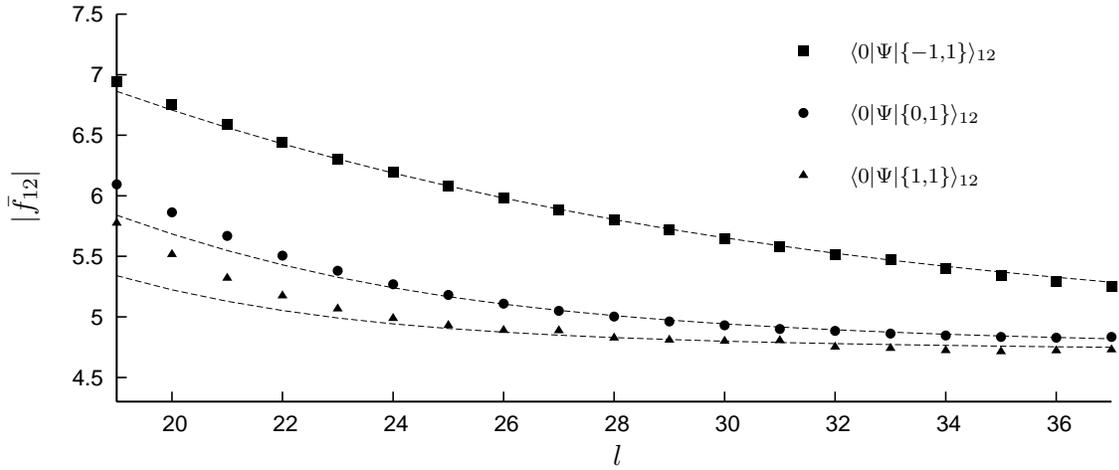}}\par\end{centering}

\caption{\label{fig:ising2pt} Two-particle form factors in the Ising model.
Dots correspond to TCSA data, while the lines show the corresponding
form factor prediction.}
\end{figure}

In this case, there is some further subtlety to be solved before proceeding
to the numerical comparison. Namely, there are spin-$0$ states which
are parity reflections of each other, but are degenerate according
to the Bethe-Yang equations. An example is the state $|\{1,-1\}\rangle_{12}$
in figure \ref{fig:ising2pt} (b), which is degenerate with $|\{-1,1\}\rangle_{12}$
to all orders in $1/L$. In general the degeneracy of these states
is lifted by residual finite size effects (more precisely by quantum
mechanical tunneling -- a detailed discussion of this mechanism was
given in the framework of the $k$-folded sine-Gordon model in \cite{kfolded}).
Since the finite volume spectrum is parity symmetric, the TCSA eigenvectors
correspond to the states\[
|\{1,-1\}\rangle_{12,\, L}^{\pm}=\frac{1}{\sqrt{2}}\left(|\{1,-1\}\rangle_{12,\, L}\pm|\{-1,1\}\rangle_{12,\, L}\right)\]
Because the Hilbert space inner product is positive definite, the
TCSA eigenvectors $|\{1,-1\}\rangle_{12}^{\pm}$ can be chosen orthonormal
and the problem can be resolved by calculating the form factor matrix
element using the two-particle state vectors\[
\frac{1}{\sqrt{2}}\left(|\{1,-1\}\rangle_{12,\, L}^{+}\pm|\{1,-1\}\rangle_{12,\, L}^{-}\right)\]
Because the Ising spectrum is much more complicated than that of the
scaling Lee-Yang model (and truncation errors are larger as well),
we only identified two-particle states containing two copies of $A_{1}$
, or an $A_{1}$ and an $A_{2}$. The numerical results are plotted
in figures \ref{fig:ising2pt} (a) and (b), respectively. The finite
volume form factor functions of the operator $\Psi$ (\ref{eq:psidef})
are defined as\[
\bar{f}_{11}\left(l\right)_{I_{1}I_{2}}=\sqrt{\rho_{11}(\tilde{\theta}_{1}(l),\tilde{\theta}_{2}(l))}\langle0|\Psi|\{ I_{1},I_{2}\}\rangle_{11}\]
where\begin{eqnarray*}
 &  & l\sinh\tilde{\theta}_{1}+\delta_{11}(\tilde{\theta}_{1}-\tilde{\theta}_{2})=2\pi I_{1}\\
 &  & l\sinh\tilde{\theta}_{2}+\delta_{11}(\tilde{\theta}_{2}-\tilde{\theta}_{1})=2\pi I_{2}\\
 &  & \rho_{11}(\theta_{1},\theta_{2})=l^{2}\cosh\theta_{1}\cosh\theta_{2}+l\cosh\theta_{1}\varphi_{11}(\theta_{2}-\theta_{1})+l\cosh\theta_{2}\varphi_{11}(\theta_{1}-\theta_{2})\\
 &  & \varphi_{11}(\theta)=\frac{d\delta_{11}(\theta)}{d\theta}\end{eqnarray*}
and\[
\bar{f}_{12}\left(l\right)_{I_{1}I_{2}}=\sqrt{\rho_{11}(\tilde{\theta}_{1},\tilde{\theta}_{2})}\langle0|\Psi|\{ I_{1},I_{2}\}\rangle_{12}\]
with\begin{eqnarray*}
 &  & l\sinh\tilde{\theta}_{1}+\delta_{12}(\tilde{\theta}_{1}-\tilde{\theta}_{2})=2\pi I_{1}\\
 &  & \frac{m_{2}}{m_{1}}l\sinh\tilde{\theta}_{2}+\delta_{12}(\tilde{\theta}_{2}-\tilde{\theta}_{1})=2\pi I_{2}\\
 &  & \rho_{12}(\theta_{1},\theta_{2})=\frac{m_{2}}{m_{1}}l^{2}\cosh\theta_{1}\cosh\theta_{2}+l\cosh\theta_{1}\varphi_{12}(\theta_{2}-\theta_{1})+\frac{m_{2}}{m_{1}}l\cosh\theta_{2}\varphi_{12}(\theta_{1}-\theta_{2})\\
 &  & \varphi_{12}(\theta)=\frac{d\delta_{12}(\theta)}{d\theta}\end{eqnarray*}
and are compared against the form factor functions\[
F_{2}^{\Psi}(\tilde{\theta}_{1}(l),\tilde{\theta}_{2}(l))_{11}\]
and \[
F_{2}^{\Psi}(\tilde{\theta}_{1}(l),\tilde{\theta}_{2}(l))_{12}\]
respectively. 

Although (as we already noted) truncation errors in the Ising model
are much larger than in the Lee-Yang case, extrapolation in the cutoff
improves them by an order of magnitude compared to the evaluation
at the highest cutoff (in our case $30$). After extrapolation, deviations
in the scaling region become less than $1$\% (with a minimum of around
$10^{-3}$ in the $A_{1}A_{1}$, and $10^{-4}$ in the $A_{1}A_{2}$
case), and even better for states with nonzero total spin. As noted
in the previous subsection this means that the numerics is really
sensitive to the dependence of the particle rapidities and state density
factors on the interaction between the particles; generally the truncation
errors in the extrapolated data are about two orders of magnitude
smaller than the interaction corrections.

It is a general tendency that the agreement is better in the sectors
with nonzero spin, and the scaling region starts at smaller values
of the volume. This is easy to understand for the energy levels, since
for low-lying states nonzero spin generally means higher particle
momenta. The higher the momenta of the particles, the more the Bethe-Yang
contributions dominate over the residual finite size effects. This
is consistent with the results of Rummukainen and Gottlieb in \cite{nonzeromom}
where it was found that resonance phase shifts can be more readily
extracted from sectors with nonzero momentum; our data show that this
observation carries over to general matrix elements as well. 

\subsection{\label{sub:manypff} Many-particle form factors}

\subsubsection{Scaling Lee-Yang model}

\begin{figure}
\begin{centering}\psfrag{f3}{$|\tilde{f}_3(l)|$}\psfrag{l}{$l$}
\psfrag{tcsa101}{$\scriptstyle\langle 0|\Phi|\{ 1,0,-1\}\rangle$}
\psfrag{tcsa202}{$\scriptstyle\langle 0|\Phi|\{ 2,0,-2\}\rangle$}
\psfrag{tcsa303}{$\scriptstyle\langle 0|\Phi|\{ 3,0,-3\}\rangle$}
\psfrag{tcsa312}{$\scriptstyle\langle 0|\Phi|\{ 3,-1,-2\}\rangle$}\includegraphics[scale=1.6]{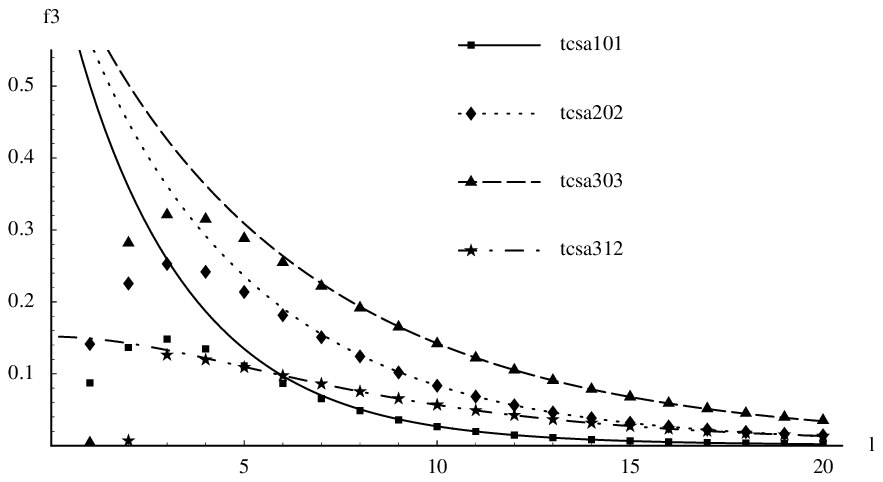}\par\end{centering}

\caption{\label{fig:lythps0}Three-particle form factors in the spin-$0$
sector. Dots correspond to TCSA data, while the lines show the corresponding
form factor prediction.}
\end{figure}

\begin{figure}
\begin{centering}\psfrag{f4}{$|\tilde{f}_4(l)|$}\psfrag{l}{$l$}
\psfrag{tcsa3113}{$\scriptstyle\langle 0|\Phi|\{ 3/2,1/2,-1/2,-3/2\}\rangle$}
\psfrag{tcsa5115}{$\scriptstyle\langle 0|\Phi|\{ 5/2,1/2,-1/2,-5/2\}\rangle$}
\psfrag{tcsa7117}{$\scriptstyle\langle 0|\Phi|\{ 7/2,1/2,-1/2,-7/2\}\rangle$}
\psfrag{tcsa7135}{$\scriptstyle\langle 0|\Phi|\{ 7/2,1/2,-3/2,-5/2\}\rangle$}\includegraphics[scale=1.6]{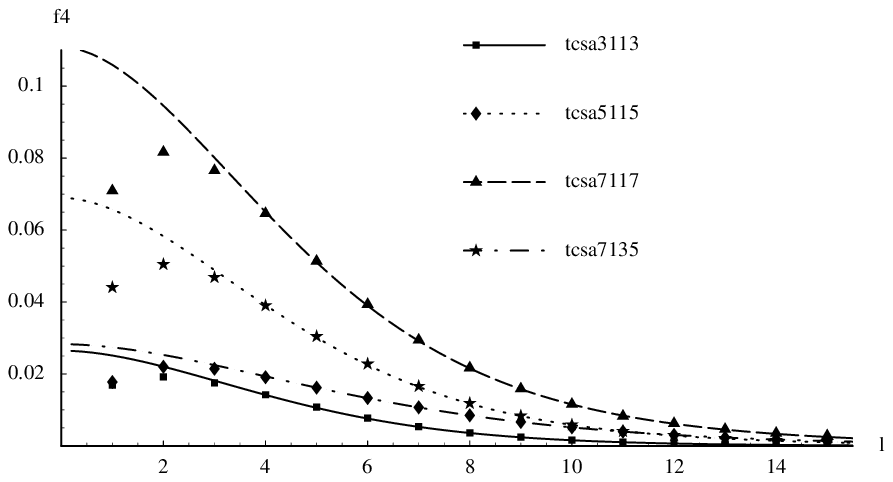}\par\end{centering}

\caption{\label{fig:lyfps0}Four-particle form factors in the spin-$0$ sector.
Dots correspond to TCSA data, while the lines show the corresponding
form factor prediction.}
\end{figure}

We also performed numerical evaluation of three and four-particle
form factors in the scaling Lee-Yang model; some of the results are
presented in figures \ref{fig:lythps0} and \ref{fig:lyfps0}, respectively.
For the sake of brevity we refrain from presenting explicit numerical
tables; we only mention that the agreement between the numerical TCSA
data and the prediction from the exact form factor solution is always
better than $10^{-3}$ in the scaling region. For better visibility
we plotted the functions\[
\tilde{f}_{k}(l)_{I_{1}\dots I_{k}}=-im^{2/5}\sqrt{\rho_{k}(\tilde{\theta}_{1},\dots,\tilde{\theta}_{k})}\langle0|\Phi|\{ I_{1},\dots,I_{k}\}\rangle_{L}\quad,\quad l=mL\]
for which relation (\ref{eq:ffrelation}) gives:\begin{equation}
\tilde{f}_{k}(l)_{I_{1}\dots I_{k}}=-im^{2/5}F_{k}^{\Phi}(\tilde{\theta}_{1},\dots,\tilde{\theta}_{k})+O(\mathrm{e}^{-l})\label{eq:lyfkprediction}\end{equation}
Due to the fact that in the Lee-Yang model there is only a single
particle species, we introduced the simplified notation $\rho_{n}$
for the $n$-particle Jacobi determinant.

The complication noted in subsection 4.2.2 for the Ising state $|\{1,-1\}\rangle_{12}$
is present in the Lee-Yang model as well. The Bethe-Yang equations
give degenerate energy values for the states $|\{ I_{1},\dots,I_{k}\}\rangle_{L}$
and $|\{-I_{k},\dots,-I_{1}\}\rangle_{L}$ (as noted before, the degeneracy
is lifted by quantum mechanical tunneling). For states with nonzero
spin this causes no problem, because these two states are in sectors
of different spin (their spins differ by a sign) and similarly there
is no difficulty when the two quantum number sets are identical, i.e.\[
\{ I_{1},\dots,I_{k}\}=\{-I_{1},\dots,-I_{k}\}\]
since then there is a single state. However, there are states in the
zero spin sector (i.e. with $\sum_{k}I_{k}=0$) for which \[
\{ I_{1},\dots,I_{k}\}\neq\{-I_{1},\dots,-I_{k}\}\]
We use two such pairs of states in our data here: the three-particle
states $|\{3,-1,-2\}\rangle_{L}$, $|\{2,1,-3\}\rangle_{L}$ and the
four-particle states $|\{7/2,1/2,-3/2,-5/2\}\rangle_{L}$, $|\{5/2,3/2,-1/2,-7/2\}\rangle_{L}$.
Again, the members of such pairs are related to each other by spatial
reflection, which is a symmetry of the exact finite-volume Hamiltonian
and therefore (supposing that the eigenvectors are orthonormal) the
finite volume eigenstates correspond to \[
|\{ I_{1},\dots,I_{k}\}\rangle_{L}^{\pm}=\frac{1}{\sqrt{2}}\left(|\{ I_{1},\dots,I_{k}\}\rangle_{L}\pm|\{-I_{k},\dots,-I_{1}\}\rangle_{L}\right)\]
and this must be taken into account when evaluating the form factor
matrix elements. In the Lee-Yang case, however, the inner product
is not positive definite (and some nonzero vectors may have zero {}``length'',
although this does not happen for TCSA eigenvectors, because they
are orthogonal to each other and the inner product is non-degenerate),
but there is a simple procedure that can be used in the general case.
Suppose the two TCSA eigenvectors corresponding to such a pair are
$v_{1}$ and $v_{2}$. Then we can define their inner product matrix
as \[
g_{ij}=v_{i}G^{(0)}v_{j}\]
using the TCSA inner product (\ref{eq:Gs}). The appropriate basis
vectors of this two-dimensional subspace, which can be identified
with $|\{ I_{1},\dots,I_{k}\}\rangle_{L}$ and $|\{-I_{k},\dots,-I_{1}\}\rangle_{L}$,
can be found by solving the two-dimensional generalized eigenvalue
problem\[
g\cdot w=\lambda P\cdot w\]
for the vector $(w_{1},w_{2})$ describing orientation in the subspace,
with \[
P=\left(\begin{array}{cc}
0 & 1\\
1 & 0\end{array}\right)\]
This procedure has the effect of rotating from the basis of parity
eigenvectors to basis vectors which are taken into each other by spatial
reflection.


\subsubsection{Ising model in a magnetic field}

\begin{figure}
\noindent \begin{centering}\psfrag{f111}{$|\tilde{f}_{111}|$}\psfrag{l}{$l$}
\psfrag{---a0a0a0}{$\scriptstyle\langle 0|\Psi|\{1,0,-1\}\rangle_{111}$}
\psfrag{---a-1a0a2}{$\scriptstyle\langle 0|\Psi|\{2,0,-1\}\rangle_{111}$}
\psfrag{---a0a1a2}{$\scriptstyle\langle 0|\Psi|\{2,1,0\}\rangle_{111}$}\includegraphics[scale=1.3]{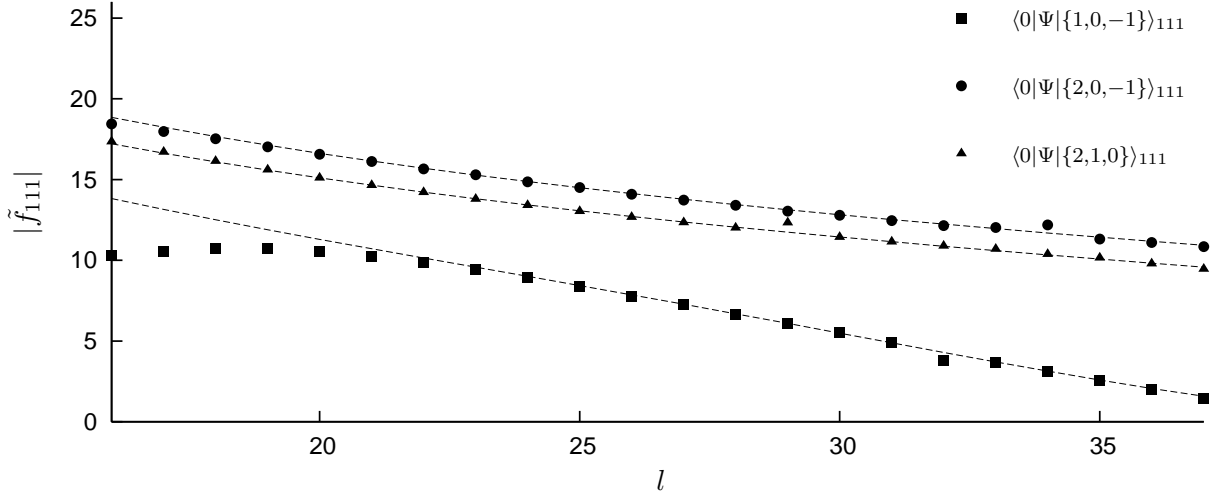}\par\end{centering}

\caption{\label{fig:ising3pt} Three-particle form factors in the Ising model.
Dots correspond to TCSA data, while the lines show the corresponding
form factor prediction.}
\end{figure}

As we already noted, it is much harder to identify%
\footnote{To identify $A_{1}A_{1}A_{1}$ states it is necessary to use at least
$e_{\mathrm{cut}}=22$ or $24$ and even then the agreement with the
Bethe-Yang prediction is still only within $20\%$, but the identification
can be made for the first few $A_{1}A_{1}A_{1}$ states using data
up to $e_{\mathrm{cut}}=30$. Truncation errors are substantially
decreased by extrapolation to $e_{\mathrm{cut}}=\infty$. %
} higher states in the Ising model due to the complexity of the spectrum,
and so we only performed an analysis of states containing three $A_{1}$
particles. We define\[
\tilde{f}_{111}\left(l\right)_{I_{1}I_{2}I_{3}}=\sqrt{\rho_{111}(\tilde{\theta}_{1}(l),\tilde{\theta}_{2}(l),\tilde{\theta}_{3}(l))}\langle0|\Psi|\{ I_{1},I_{2},I_{3}\}\rangle_{111}\]
where $\tilde{\theta}_{i}(l)$ are the solutions of the three-particle
Bethe-Yang equations in (dimensionless) volume $l$ and $\rho_{111}$
is the appropriate $3$-particle determinant. The results of the comparison
can be seen in figure \ref{fig:ising3pt}. The numerical precision
indicated for two-particle form factors at the end of subsection 4.2.2,
as well as the remarks made there on the spin dependence apply here
as well; we only wish to emphasize that for $A_{1}A_{1}A_{1}$ states
with nonzero total spin the agreement between the extrapolated TCSA
data and the form factor prediction in the optimal part of the scaling
region is within $2\times10^{-4}$. 

\section{General form factors without disconnected pieces}

Let us consider a matrix element of the form\[
\,_{j_{1}\dots j_{m}}\langle\{ I_{1}',\dots,I_{m}'\}\vert\mathcal{O}(0,0)\vert\{ I_{1},\dots,I_{n}\}\rangle_{i_{1}\dots i_{n},L}\]
Disconnected pieces are known to appear when there is at least one
particle in the state on the left which occurs in the state on the
right with exactly the same rapidity. The rapidities of particles
as a function of the volume are determined by the Bethe-Yang equations
(\ref{eq:betheyang})\[
Q_{k}(\tilde{\theta}_{1},\dots,\tilde{\theta}_{n})_{i_{1}\dots i_{n}}=m_{i_{k}}L\sinh\tilde{\theta}_{k}+\sum_{l\neq k}\delta_{i_{k}i_{l}}(\tilde{\theta}_{k}-\tilde{\theta}_{l})=2\pi I_{k}\quad,\quad k=1,\dots,n\]
and \[
Q_{k}(\tilde{\theta}_{1}',\dots,\tilde{\theta}_{m}')_{j_{1}\dots j_{m}}=m_{j_{k}}L\sinh\tilde{\theta}_{k}'+\sum_{l\neq k}\delta_{j_{k}j_{l}}(\tilde{\theta}_{k}'-\tilde{\theta}_{l}')=2\pi I_{k}'\quad,\quad k=1,\dots,m\]
Due to the presence of the scattering terms containing the phase shift
functions $\delta$, equality of two quantum numbers $I_{k}$ and
$I_{l}'$ does not mean that the two rapidities themselves are equal
in finite volume $L$. It is easy to see that there are only two cases
when exact equality of some rapidities can occur:

\begin{enumerate}
\item The two states are identical, i.e. $n=m$ and \begin{eqnarray*}
\{ j_{1}\dots j_{m}\} & = & \{ i_{1}\dots i_{n}\}\\
\{ I_{1}',\dots,I_{m}'\} & = & \{ I_{1},\dots,I_{n}\}\end{eqnarray*}
in which case all the rapidities are pairwise equal, or
\item Both states are parity symmetric states in the spin zero sector, i.e.
\begin{eqnarray*}
\{ I_{1},\dots,I_{n}\} & \equiv & \{-I_{n},\dots,-I_{1}\}\\
\{ I_{1}',\dots,I'_{m}\} & \equiv & \{-I'_{m},\dots,-I'_{1}\}\end{eqnarray*}
and the particle species labels are also compatible with the symmetry,
i.e. $i_{n+1-k}=i_{k}$ and $j_{m+1-k}=j_{k}$. Furthermore, both
states must contain one (or possibly more, in a theory with more than
one species) particle of quantum number $0$, whose rapidity is then
exactly $0$ for any value of the volume $L$ due to the symmetric
assignment of quantum numbers.
\end{enumerate}
Discussion of such matrix elements raises many interesting theoretical
considerations and is postponed to the followup paper \cite{crossing};
here we only concentrate on matrix elements for which there are no
disconnected contributions.

\subsection{Scaling Lee-Yang model}

In this model there is a single particle species, so we can introduce
the following notations:\[
f_{kn}(l)_{I_{1},\dots,I_{n}}^{I_{1}',\dots,I_{k}'}=-im^{2/5}\langle\{ I_{1}',\dots,I_{k}'\}\vert\Phi(0,0)\vert\{ I_{1},\dots,I_{n}\}\rangle_{L}\]
and also \[
\tilde{f}_{kn}(l)_{I_{1},\dots,I_{n}}^{I_{1}',\dots,I_{k}'}=-im^{2/5}\sqrt{\rho_{k}(\tilde{\theta}_{1}',\dots,\tilde{\theta}_{k}')}\sqrt{\rho_{n}(\tilde{\theta}_{1},\dots,\tilde{\theta}_{n})}\langle\{ I_{1}',\dots,I_{k}'\}\vert\Phi(0,0)\vert\{ I_{1},\dots,I_{n}\}\rangle_{L}\]
for which relation (\ref{eq:genffrelation}) yields\begin{eqnarray}
f_{kn}(l)_{I_{1},\dots,I_{n}}^{I_{1}',\dots,I_{k}'} & = & -im^{2/5}\frac{F_{k+n}^{\Phi}(\tilde{\theta}_{k}'+i\pi,\dots,\tilde{\theta}_{1}'+i\pi,\tilde{\theta}_{1},\dots,\tilde{\theta}_{n})}{\sqrt{\rho_{n}(\tilde{\theta}_{1},\dots,\tilde{\theta}_{n})\rho_{k}(\tilde{\theta}_{1}',\dots,\tilde{\theta}_{m}')}}+O(\mathrm{e}^{-l})\nonumber \\
\tilde{f}_{kn}(l)_{I_{1},\dots,I_{n}}^{I_{1}',\dots,I_{k}'} & = & -im^{2/5}F_{k+n}^{\Phi}(\tilde{\theta}_{k}'+i\pi,\dots,\tilde{\theta}_{1}'+i\pi,\tilde{\theta}_{1},\dots,\tilde{\theta}_{n})+O(\mathrm{e}^{-l})\label{eq:fmnrels}\end{eqnarray}
For the plots we chose to display $f$ or $\tilde{f}$ depending on
which one gives a better visual picture. The numerical results shown
here are just a fraction of the ones we actually obtained, but all
of them show an agreement with precision $10^{-4}-10^{-3}$ in the
scaling region (the volume range corresponding to the scaling region
typically varies depending on the matrix element considered due to
variation in the residual finite size corrections and truncation effects).

\begin{figure}
\noindent \begin{centering}\psfrag{f11}{$|\tilde{f}_{11}(l)|$}\psfrag{l}{$l$}
\psfrag{tcsa01}{$\scriptstyle\langle\{ 0\}|\Phi|\{ 1\}\rangle$}
\psfrag{tcsa02}{$\scriptstyle\langle\{ 0\}|\Phi|\{ 2\}\rangle$}
\psfrag{tcsa1m1}{$\scriptstyle\langle\{ -1\}|\Phi|\{ 1\}\rangle$}
\psfrag{tcsa12}{$\scriptstyle\langle\{ 1\}|\Phi|\{ 2\}\rangle$}
\psfrag{tcsam12}{$\scriptstyle\langle\{ -1\}|\Phi|\{ 2\}\rangle$}
\includegraphics[scale=1.4]{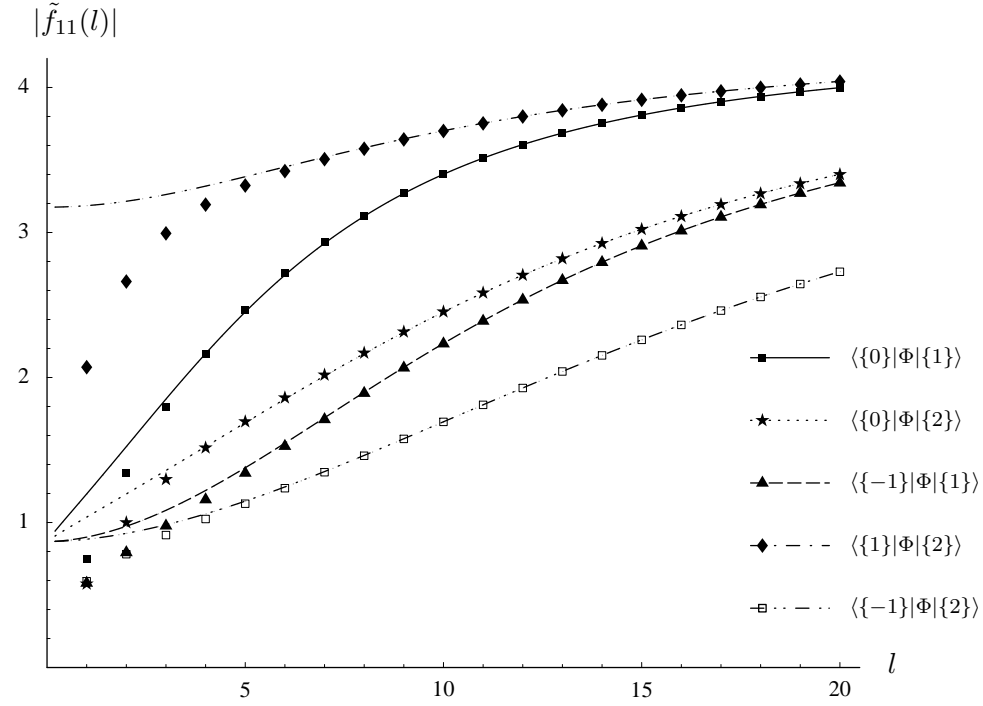}\par\end{centering}

\caption{\label{fig:ly1r1r} One-particle--one-particle form factors in Lee-Yang
model. Dots correspond to TCSA data, while the lines show the corresponding
form factor prediction.}
\end{figure}

\begin{figure}
\noindent \begin{raggedright}\psfrag{f12}{$|{f}_{12}(l)|$}\psfrag{l}{$l$}
\psfrag{tcsa01m1}{$\scriptstyle\langle\{ 0\}|\Phi|\{ 1/2,-1/2\}\rangle$}
\psfrag{tcsa03m3}{$\scriptstyle\langle\{ 0\}|\Phi|\{ 3/2,-3/2\}\rangle$}
\psfrag{tcsa01m3}{$\scriptstyle\langle\{ 0\}|\Phi|\{ 1/2,-3/2\}\rangle$}
\psfrag{tcsa21m3}{$\scriptstyle\langle\{ 2\}|\Phi|\{ 1/2,-3/2\}\rangle$}

\includegraphics[scale=1.4]{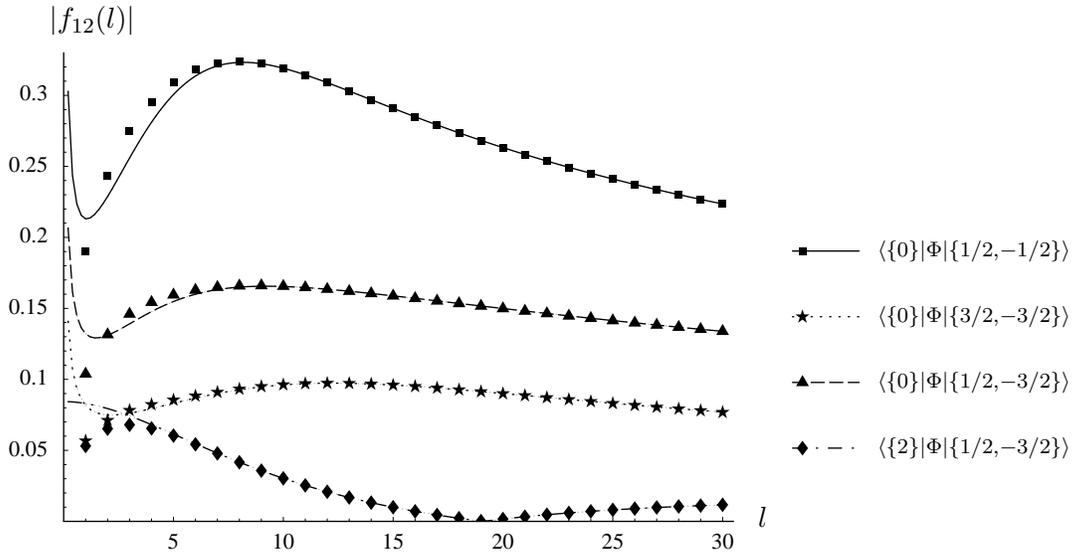}\par\end{raggedright}

\caption{\label{fig:ly1r2r} One-particle--two-particle form factors in Lee-Yang
model. Dots correspond to TCSA data, while the lines show the corresponding
form factor prediction.}
\end{figure}

The simplest cases involve one and two-particle states: the one-particle--one-particle
data in figure \ref{fig:ly1r1r} actually test the two-particle form
factor $F_{2}^{\Phi}$, while the one-particle--two-particle plot
\ref{fig:ly1r2r} corresponds to $F_{3}^{\Phi}$ (we obtained similar
results on $F_{4}^{\Phi}$ using matrix elements $f_{22}$). Note
that in contrast to the comparisons performed in subsections 4.2 and
4.3, these cases involve the form factor solutions (\ref{eq:lyff})
at complex values of the rapidities. In general, all tests performed
with TCSA can test form factors at rapidity arguments with imaginary
parts $0$ or $\pi$, which are the only parts of the complex rapidity
plane where form factors eventually correspond to physical matrix
elements. 


\begin{figure}
\noindent \begin{raggedright}\psfrag{f13}{$|{f}_{13}(l)|$}\psfrag{l}{$l$}
\psfrag{tcsa110m1}{$\scriptstyle\langle\{ 1\}|\Phi|\{ 1,0,-1\}\rangle$}
\psfrag{tcsa120m2}{$\scriptstyle\langle\{ 1\}|\Phi|\{ 2,0,-2\}\rangle$}
\psfrag{tcsa03m1m2}{$\scriptstyle\langle\{ 0\}|\Phi|\{ 3,-1,-2\}\rangle$}
\psfrag{tcsa13m1m2}{$\scriptstyle\langle\{ 1\}|\Phi|\{ 3,-1,-2\}\rangle$}
\psfrag{tcsa1m312}{$\scriptstyle\langle\{ 1\}|\Phi|\{ -3,1,2\}\rangle$}
\includegraphics[scale=1.4]{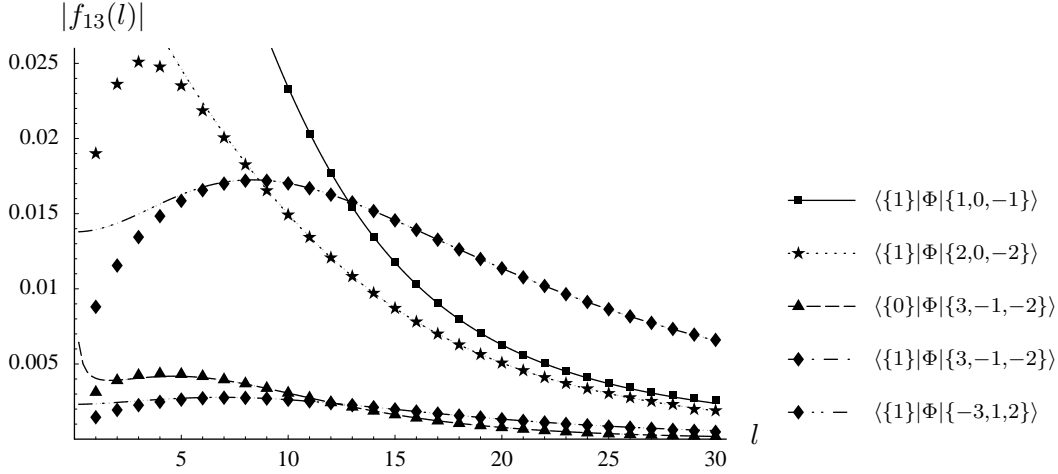}\par\end{raggedright}

\caption{\label{fig:ly1r3r} One-particle--three-particle form factors in
Lee-Yang model. Dots correspond to TCSA data, while the lines show
the corresponding form factor prediction.}
\end{figure}

\begin{figure}
\noindent \begin{raggedright}\psfrag{f14}{$|{f}_{14}(l)|$}\psfrag{l}{$l$}
\psfrag{tcsa01}{$\scriptstyle\langle\{ 0\}|\Phi|\{ 3/2,1/2,-1/2,-3/2\}\rangle$}
\psfrag{tcsa02}{$\scriptstyle\langle\{ 0\}|\Phi|\{ 5/2,1/2,-1/2,-5/2\}\rangle$}
\psfrag{tcsa12}{$\scriptstyle\langle\{ 1\}|\Phi|\{ 3/2,1/2,-1/2,-3/2\}\rangle$}
\psfrag{tcsa14}{$\scriptstyle\langle\{ 1\}|\Phi|\{ 5/2,3/2,-1/2,-7/2\}\rangle$}
\psfrag{tcsa15}{$\scriptstyle\langle\{ 1\}|\Phi|\{ 7/2,1/2,-3/2,-5/2\}\rangle$}
\includegraphics[scale=1.4]{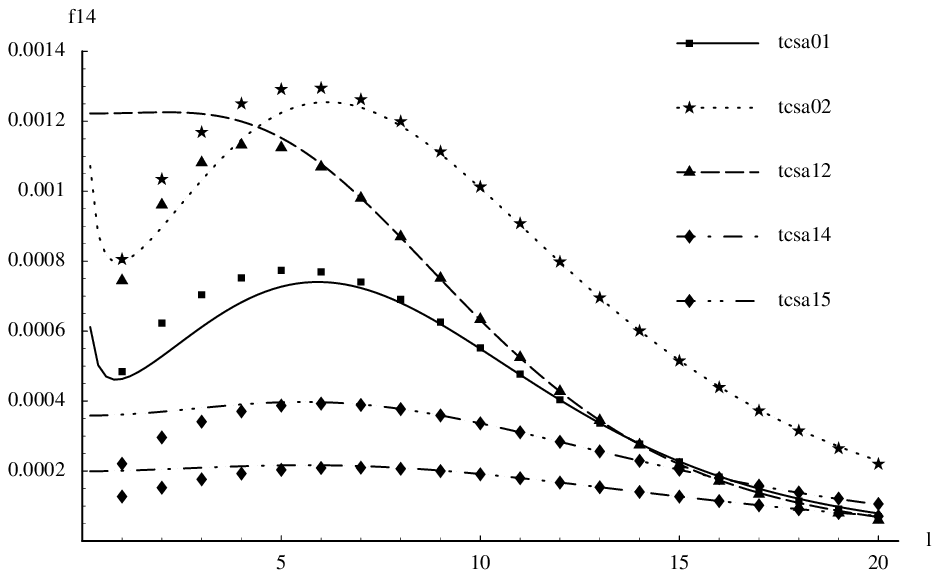}\par\end{raggedright}

\caption{\label{fig:ly1r4r} One-particle--four-particle form factors in Lee-Yang
model. Dots correspond to TCSA data, while the lines show the corresponding
form factor prediction.}
\end{figure}

\begin{figure}
\noindent \begin{raggedright}\psfrag{f23}{$|{f}_{23}(l)|$}\psfrag{l}{$l$}
\psfrag{tcsa11}{$\scriptstyle\langle\{ 1/2,-1/2\}|\Phi|\{ 1,0,-1\}\rangle$}
\psfrag{tcsa12}{$\scriptstyle\langle\{ 3/2,-3/2\}|\Phi|\{ 1,0,-1\}\rangle$}
\psfrag{tcsa21}{$\scriptstyle\langle\{ 1/2,-1/2\}|\Phi|\{ 2,0,-2\}\rangle$}
\psfrag{tcsas14}{$\scriptstyle\langle\{ 3/2,-1/2\}|\Phi|\{ 3,-1,-2\}\rangle$}
\psfrag{tcsas15}{$\scriptstyle\langle\{ 3/2,-1/2\}|\Phi|\{ -3,1,2\}\rangle$}
\includegraphics[scale=1.4]{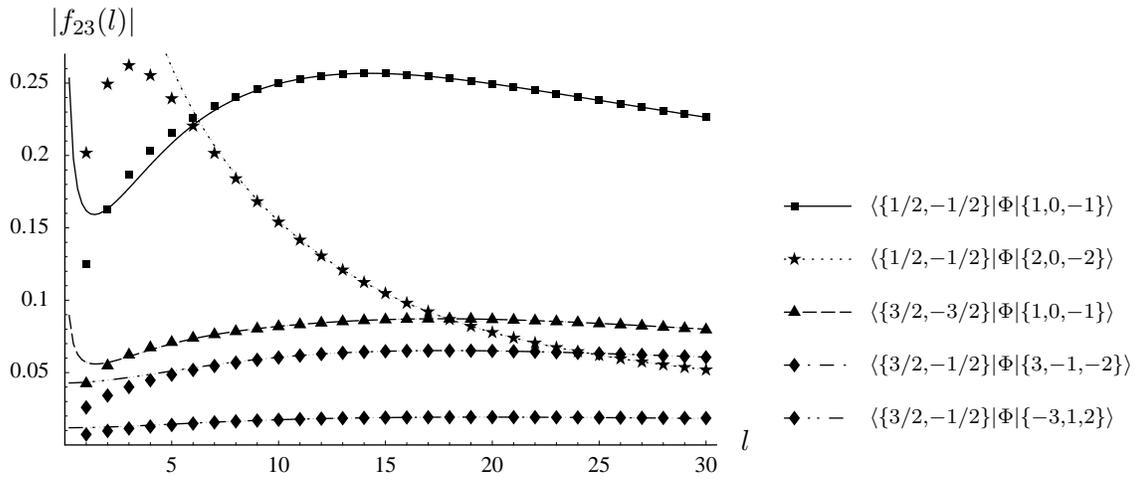}\par\end{raggedright}

\caption{\label{fig:ly2r3r} Two-particle--three-particle form factors in
Lee-Yang model. Dots correspond to TCSA data, while the lines show
the corresponding form factor prediction.}
\end{figure}

One-particle--three-particle and one-particle--four-particle matrix
elements $f_{13}$ and $f_{14}$ contribute another piece of useful
information. We recall that there are pairs of parity-related states
in the spin-0 factors which we cannot distinguish in terms of their
elementary form factors. In subsection 4.3 we showed the example of
the three-particle states\[
|\{3,-1,-2\}\rangle_{L}\mbox{ and }|\{2,1,-3\}\rangle_{L}\]
and the four-particle states\[
|\{7/2,1/2,-3/2,-5/2\}\rangle_{L}\mbox{ and }|\{5/2,3/2,-1/2,-7/2\}\rangle_{L}\]
In fact it is only true that they cannot be distinguished if the left
state is parity-invariant. However, using a one-particle state of
nonzero spin on the left it is possible to distinguish and appropriately
label the two states, as shown in figures \ref{fig:ly1r3r} and \ref{fig:ly1r4r}.
This can also be done using matrix elements with two-particle states
of nonzero spin: the two-particle--three-particle case $f_{23}$ is
shown in \ref{fig:ly2r3r} (similar results were obtained for $f_{24}$).
Examining the data in detail shows that the identifications provided
using different states on the left are all consistent with each other. 


\begin{figure}
\noindent \begin{raggedright}\psfrag{f3344}{$|{f}_{33}(l)|$ or $|{f}_{44}(l)|$}\psfrag{l}{$l$}
\psfrag{tcsa3314}{$\scriptstyle\langle\{ 1,0,-1\}|\Phi|\{ 3,-1,-2\}\rangle$}
\psfrag{tcsa3324}{$\scriptstyle\langle\{ 1,0,-1\}|\Phi|\{ 3,-1,-2\}\rangle$}
\psfrag{tcsa3334}{$\scriptstyle\langle\{ 1,0,-1\}|\Phi|\{ 3,-1,-2\}\rangle$}
\psfrag{tcsas4412}{$\scriptstyle\langle\{ \frac{\scriptstyle 3}{\scriptstyle 2},\frac{\scriptstyle 1}{\scriptstyle 2},-\frac{\scriptstyle 1}{\scriptstyle 2},-\frac{\scriptstyle 3}{\scriptstyle 2}\}|\Phi|\{\frac{\scriptstyle 5}{\scriptstyle 2},\frac{\scriptstyle 1}{\scriptstyle 2},-\frac{\scriptstyle 1}{\scriptstyle 2},-\frac{\scriptstyle 5}{\scriptstyle 2}\}\rangle$}
\psfrag{tcsas4413}{$\scriptstyle\langle\{ \frac{\scriptstyle 3}{\scriptstyle 2},\frac{\scriptstyle 1}{\scriptstyle 2},-\frac{\scriptstyle 1}{\scriptstyle 2},-\frac{\scriptstyle 3}{\scriptstyle 2}\}|\Phi|\{ \frac{\scriptstyle 7}{\scriptstyle 2},\frac{\scriptstyle 1}{\scriptstyle 2},-\frac{\scriptstyle 1}{\scriptstyle 2},-\frac{\scriptstyle 7}{\scriptstyle 2}\}\rangle$}
\includegraphics[scale=1.4]{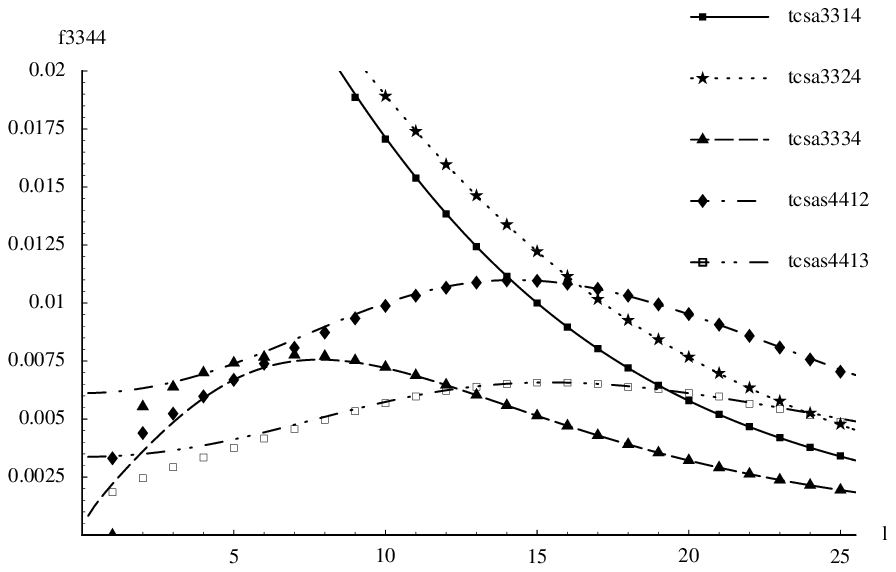}\par\end{raggedright}

\caption{\label{fig:ly3r3rand4r4r} Three-particle--three-particle and four-particle--four-particle
form factors in Lee-Yang model. Dots correspond to TCSA data, while
the lines show the corresponding form factor prediction.}
\end{figure}

\begin{figure}
\noindent \begin{raggedright}\psfrag{f34}{$|{f}_{34}(l)|$}\psfrag{l}{$l$}
\psfrag{tcsa11}{$\scriptstyle\langle\{ 1,0,-1\}|\Phi|\{ \frac{\scriptstyle 3}{\scriptstyle 2},\frac{\scriptstyle 1}{\scriptstyle 2},-\frac{\scriptstyle 1}{\scriptstyle 2},-\frac{\scriptstyle 3}{\scriptstyle 2}\}\rangle$}
\psfrag{tcsa12}{$\scriptstyle\langle\{ 1,0,-1\}|\Phi|\{ \frac{\scriptstyle 5}{\scriptstyle 2},\frac{\scriptstyle 1}{\scriptstyle 2},-\frac{\scriptstyle 1}{\scriptstyle 2},-\frac{\scriptstyle 5}{\scriptstyle 2}\}\rangle$}
\psfrag{tcsa22}{$\scriptstyle\langle\{ 2,0,-2\}|\Phi|\{ \frac{\scriptstyle 5}{\scriptstyle 2},\frac{\scriptstyle 1}{\scriptstyle 2},-\frac{\scriptstyle 1}{\scriptstyle 2},-\frac{\scriptstyle 5}{\scriptstyle 2}\}\rangle$}
\psfrag{tcsa44}{$\scriptstyle\langle\{ 3,-1,-2\}|\Phi|\{\frac{\scriptstyle 7}{\scriptstyle 2},\frac{\scriptstyle 1}{\scriptstyle 2},-\frac{\scriptstyle 3}{\scriptstyle 2},-\frac{\scriptstyle 5}{\scriptstyle 2}\}\rangle$}
\psfrag{tcsa45}{$\scriptstyle\langle\{ 2,1,-3\}|\Phi|\{ \frac{\scriptstyle 7}{\scriptstyle 2},\frac{\scriptstyle 1}{\scriptstyle 2},-\frac{\scriptstyle 3}{\scriptstyle 2},-\frac{\scriptstyle 5}{\scriptstyle 2}\}\rangle$}
\includegraphics[scale=1.4]{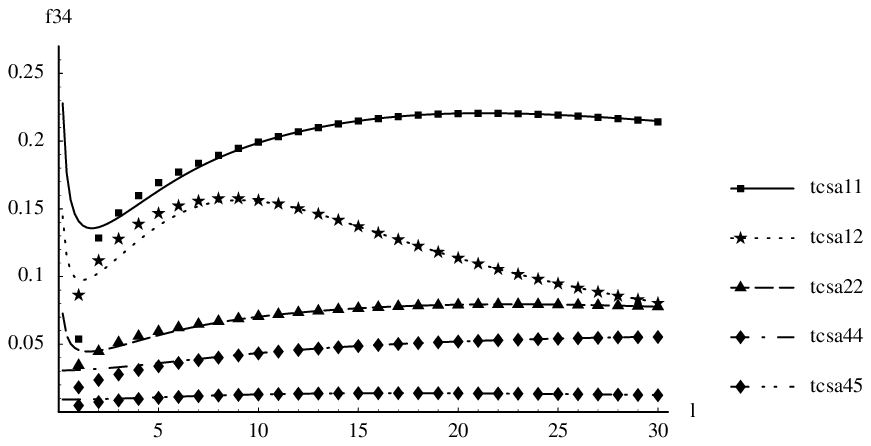}\par\end{raggedright}

\caption{\label{fig:ly3r4r} Three-particle--four-particle form factors in
Lee-Yang model. Dots correspond to TCSA data, while the lines show
the corresponding form factor prediction.}
\end{figure}

It is also interesting to note that the $f_{14}$ (figure \ref{fig:ly1r4r})
and $f_{23}$ data (figure \ref{fig:ly2r3r}) provide a test for the
five-particle form factor solutions $F_{5}$. This is important since
it is progressively harder to identify many-particle states in the
TCSA spectrum for two reasons. First, the spectrum itself becomes
more and more dense as we look for higher levels; second, the truncation
errors grow as well. Both of these make the identification of the
energy levels by comparison with the predictions of the Bethe-Yang
equations more difficult; in the Lee-Yang case we stopped at four-particle
levels. However, using general matrix elements and the relations (\ref{eq:fmnrels})
we can even get data for form factors up to $8$ particles, a sample
of which is shown in figures \ref{fig:ly3r3rand4r4r} ($f_{33}$ and
$f_{44}$, corresponding to $6$ and $8$ particle form-factors) and
\ref{fig:ly3r4r} ($f_{34}$ which corresponds to $7$ particle form
factors).


\subsection{Ising model in magnetic field}

In the case of the Ising model, we define the functions\[
\tilde{f}_{j_{1}\dots j_{m};i_{1}\dots i_{n}}(l)=\sqrt{\rho_{i_{1}\dots i_{n}}(\tilde{\theta}_{1},\dots,\tilde{\theta}_{n})\rho_{j_{1}\dots j_{m}}(\tilde{\theta}_{1}',\dots,\tilde{\theta}_{m}')}\times\,_{j_{1}\dots j_{m}}\langle\{ I_{1}',\dots,I_{m}'\}\vert\Psi\vert\{ I_{1},\dots,I_{n}\}\rangle_{i_{1}\dots i_{n},L}\]
which are compared against form factors\[
F_{m+n}^{\Psi}(\tilde{\theta}_{m}'+i\pi,\dots,\tilde{\theta}_{1}'+i\pi,\tilde{\theta}_{1},\dots,\tilde{\theta}_{n})_{j_{m}\dots j_{1}i_{1}\dots i_{n}}\]
where $\tilde{\theta}_{i}$ and $\tilde{\theta}_{j}'$ denote the
rapidities obtained as solutions of the appropriate Bethe-Yang equations
at the given value of the volume. We chose states for which the necessary
form factor solution was already known (and given in \cite{isingff})
i.e. we did not construct new form factor solutions ourselves. %
\begin{figure}
\noindent \begin{centering}\psfrag{f11}{$|\tilde{f}_{1;1}|$}\psfrag{l}{$l$}
\psfrag{a-1---a1}{$\scriptstyle{}_{1}\langle \{-1\}|\Psi|\{1\}\rangle_{1}$}
\psfrag{a-2---a2}{$\scriptstyle{}_{1}\langle \{-2\}|\Psi|\{2\}\rangle_{1}$}
\psfrag{a-3---a3}{$\scriptstyle{}_{1}\langle \{-3\}|\Psi|\{3\}\rangle_{1}$}\subfigure[$A_1-A_1$ matrix elements]{\includegraphics[scale=1.2]{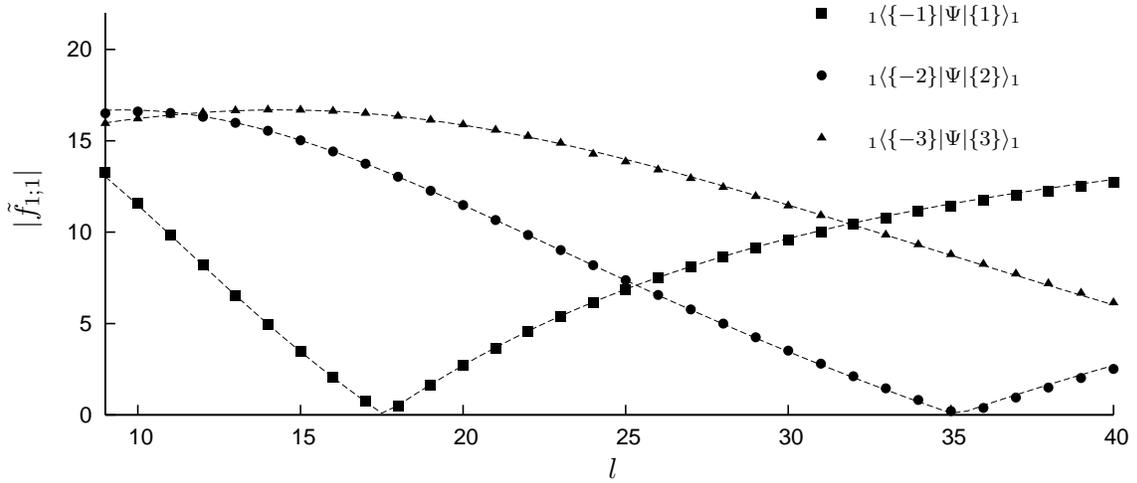}}\\
\psfrag{f12}{$|\tilde{f}_{1;2}|$}\psfrag{l}{$l$}
\psfrag{a0---b0}{$\scriptstyle{}_{1}\langle \{0\}|\Psi|\{0\}\rangle_{2}$}
\psfrag{a-1---b2}{$\scriptstyle{}_{1}\langle \{-1\}|\Psi|\{2\}\rangle_{2}$}
\psfrag{a-2---b2}{$\scriptstyle{}_{1}\langle \{-2\}|\Psi|\{2\}\rangle_{2}$}\subfigure[$A_1-A_2$ matrix elements]{\includegraphics[scale=1.2]{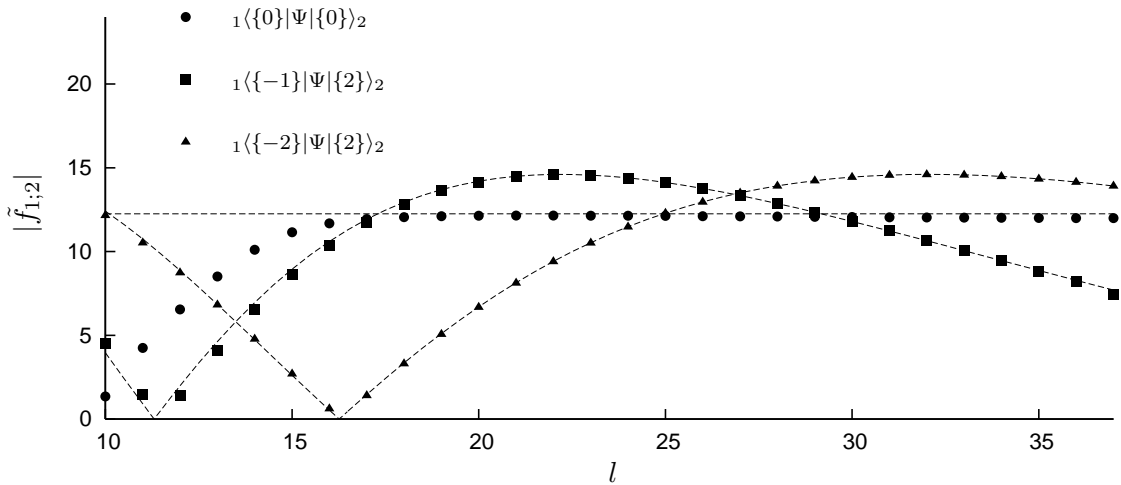}}\par\end{centering}

\caption{\label{fig:ising1p1p} One-particle--one-particle form factors in
the Ising model. Dots correspond to TCSA data, while the lines show
the corresponding form factor prediction.}
\end{figure}

\begin{figure}
\noindent \begin{centering}\psfrag{f111}{$|\tilde{f}_{1;11}|$}\psfrag{l}{$l$}
\psfrag{a0---a-0.5a0.5}{$\scriptstyle{}_{1}\langle \{0\}|\Psi|\{1/2,-1/2\}\rangle_{11}$}
\psfrag{a1---a-0.5a0.5}{$\scriptstyle{}_{1}\langle \{1\}|\Psi|\{1/2,-1/2\}\rangle_{11}$}
\psfrag{a2---a-0.5a0.5}{$\scriptstyle{}_{1}\langle \{2\}|\Psi|\{1/2,-1/2\}\rangle_{11}$}
\psfrag{a-1---a-0.5a1.5}{$\scriptstyle{}_{1}\langle \{-1\}|\Psi|\{3/2,-1/2\}\rangle_{11}$}\subfigure[$A_1-A_1A_1$ matrix elements]{\includegraphics[scale=1.3]{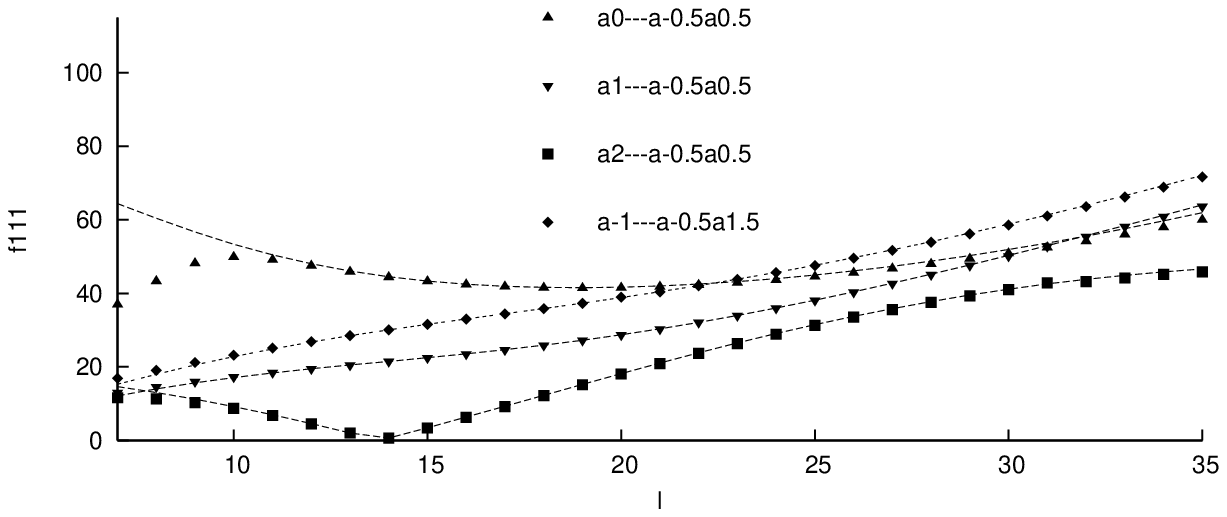}}\\
\psfrag{f112}{$|\tilde{f}_{1;12}|$}\psfrag{l}{$l$}
\psfrag{a0---a0b1}{$\scriptstyle{}_{1}\langle \{0\}|\Psi|\{1,0\}\rangle_{21}$}
\psfrag{a-1--a0b1}{$\scriptstyle{}_{1}\langle \{-1\}|\Psi|\{1,0\}\rangle_{21}$}
\psfrag{a0---a-1b1}{$\scriptstyle{}_{1}\langle \{0\}|\Psi|\{-1,1\}\rangle_{12}$}\subfigure[$A_1-A_1A_2$ matrix elements]{\includegraphics[scale=1.3]{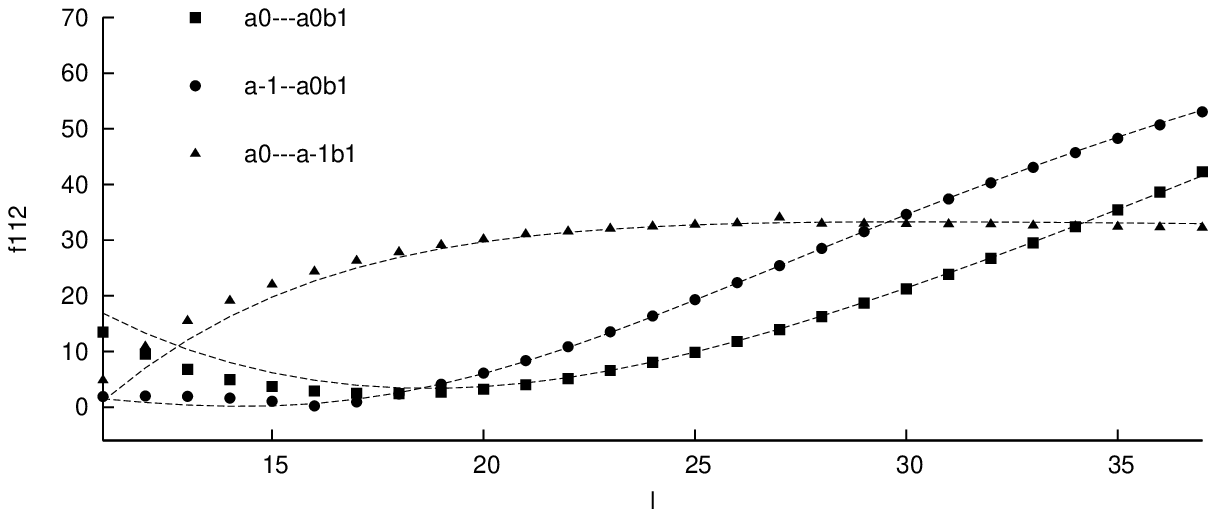}}\par\end{centering}

\caption{\label{fig:ising1p2p} One-particle--two-particle form factors in
the Ising model. Dots correspond to TCSA data, while the lines show
the corresponding form factor prediction.}
\end{figure}

One-particle--one-particle form factors are shown in figure \ref{fig:ising1p1p};
these provide another numerical test for the two-particle form factors
examined previously in subsection 4.2.2. One-particle--two-particle
form factors, besides testing again the three-particle form factor
$A_{1}A_{1}A_{1}$ (figure \ref{fig:ising1p2p} (a)) also provide
information on $A_{1}A_{1}A_{2}$ (figure \ref{fig:ising1p2p} (b)). 

\begin{figure}
\noindent \begin{centering}\psfrag{f1111}{$|\tilde{f}_{1;111}|$}\psfrag{l}{$l$}
\psfrag{a0---a-1a0a2}{$\scriptstyle{}_{1}\langle \{0\}|\Psi|\{2,0,-1\}\rangle_{111}$}
\psfrag{a-1---a-1a0a2}{$\scriptstyle{}_{1}\langle \{-1\}|\Psi|\{2,0,-1\}\rangle_{111}$}
\psfrag{a-1---a-1a1a2}{$\scriptstyle{}_{1}\langle \{-1\}|\Psi|\{2,1,-1\}\rangle_{111}$}\subfigure[$A_1-A_1A_1A_1$ matrix elements]{\includegraphics[scale=1.2]{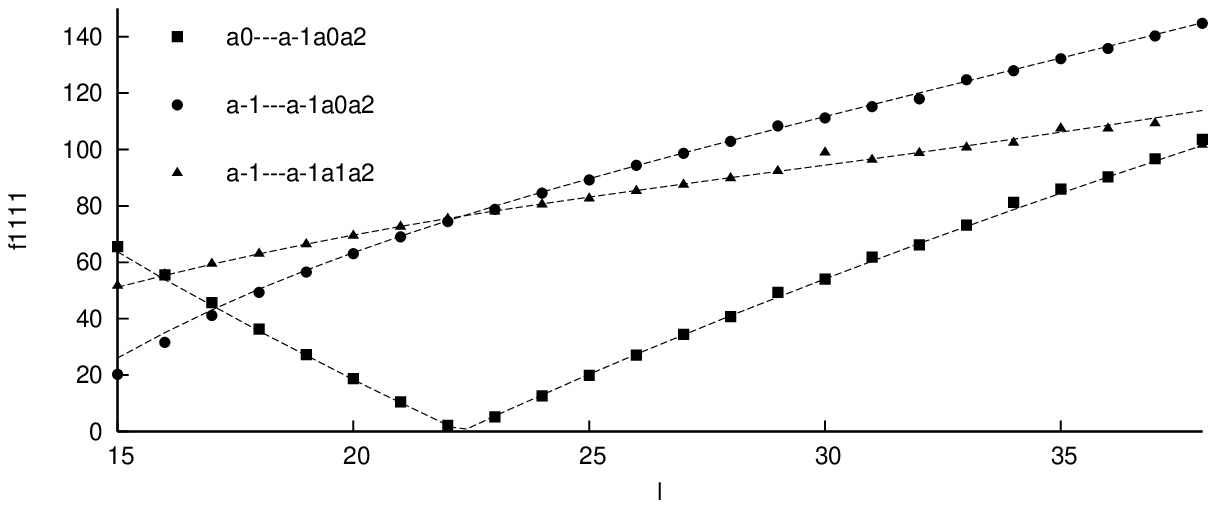}}\\
\psfrag{f1111}{$|\tilde{f}_{11;11}|$}\psfrag{l}{$l$}
\psfrag{a-1.5a0.5---a-0.5a1.5}{$\scriptstyle{}_{11}\langle \{1/2,-3/2\}|\Psi|\{3/2,-1/2\}\rangle_{11}$}
\psfrag{a-0.5a-1.5---a0.5a1.5}{$\scriptstyle{}_{11}\langle \{-1/2,-3/2\}|\Psi|\{3/2,1/2\}\rangle_{11}$}
\psfrag{a-0.5a1.5---a0.5a1.5}{$\scriptstyle{}_{11}\langle \{-1/2,3/2\}|\Psi|\{3/2,1/2\}\rangle_{11}$}\subfigure[$A_1A_1-A_1A_1$ matrix elements]{\includegraphics[scale=1.2]{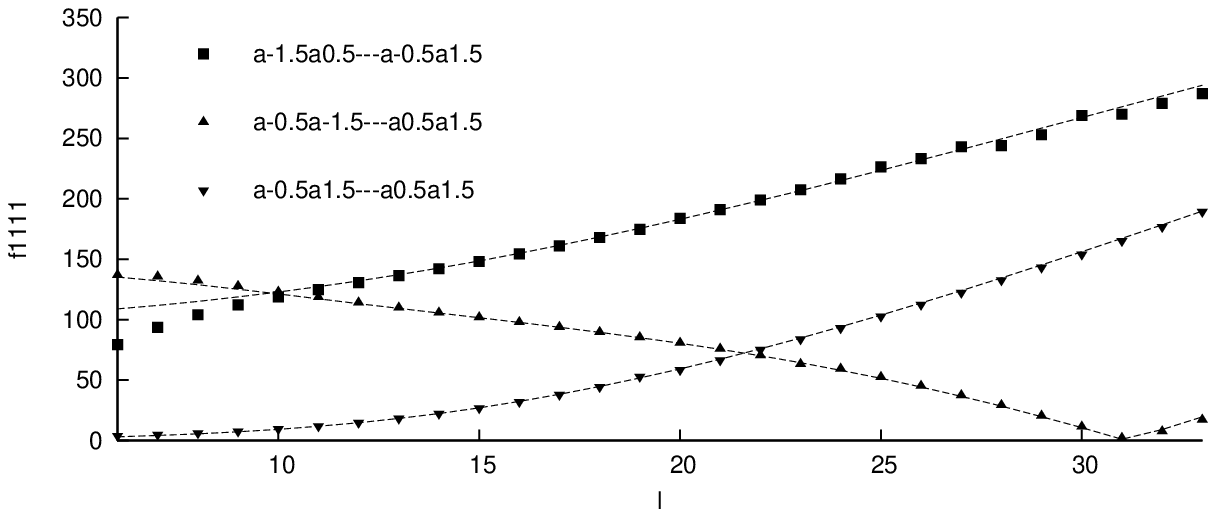}}\par\end{centering}

\caption{\label{fig:ising1p3pand2p2p}One-particle--three-particle and two-particle--two-particle
form factors in the Ising model. Dots correspond to TCSA data, while
the lines show the corresponding form factor prediction.}
\end{figure}

Finally, one-particle--three-particle and two-particle--two-particle
matrix elements can be compared to the $A_{1}A_{1}A_{1}A_{1}$ form
factor, which again shows that by considering general matrix elements
we can go substantially higher in the form factor tree than using
only elementary form factors. 

We remark that the cusps on the horizontal axis in the form factor
plots correspond to zeros where the form factors change sign; they
are artifacts introduced by taking the absolute value of the matrix
elements. The pattern of numerical deviations between TCSA data and
exact form factor predictions is fully consistent with the discussion
in the closing paragraphs of subsections 4.2.2 and 4.3.2. The deviations
in the scaling region are around $1$\% on average, with agreement
of the order of $10^{-3}$ in the optimal range.

\section{Conclusions}

In this work we gave an expression for the matrix elements of local
operators between general multi-particle states in finite volume which
is valid to any order in the expansion in the inverse of the volume
$1/L$. It was shown in section 2.2 using a very general argument
that all the remaining volume dependence is non-analytic in $1/L$
(it is given by residual finite size effects vanishing exponentially
with increasing volume). It is also clear that the derivation itself
does not depend on integrability, neither it is restricted to $1+1$
dimensional field theories and therefore relations (\ref{eq:ffrelation})
and (\ref{eq:genffrelation}) can be extended to general quantum field
theories (substituting the rapidities with appropriate kinematical
parametrization and the $\rho_{n}$ with the proper state densities),
with the only condition that their spectrum of excitations must possess
a mass gap. 

$1+1$ dimensional integrable field theories are special in the respect
that multi-particle states in finite volume can be described using
the Bethe-Yang equations (\ref{eq:betheyang}) and so the $n$-particle
state density $\rho_{n}$ can be obtained in the general closed form
(\ref{eq:byjacobian}). Another important feature is that there are
exact results for matrix elements of local operators in infinite volume
which can be obtained from the form factor bootstrap briefly reviewed
in section 2.1. Therefore they are ideal toy models to test ideas
about finite size corrections. Such an approach is also interesting
due to a fundamental property of the bootstrap, namely that it is
only indirectly related to the actual Lagrangian (or Hamiltonian)
field theory. As we discussed in the introduction, testing the conjectured
form factors against field theory usually involves calculating two-point
functions using spectral representations, or sum rules derived from
such expansions; however, direct non-perturbative comparison of the
actual form factors to matrix elements computed from the field theory
have been very restricted so far. 

Using TCSA we were able to give an extensive and direct numerical
comparison between bootstrap results for form factors and matrix elements
evaluated non-perturbatively. One of the advantages is that we can
compare matrix elements directly, without using any proxy (such as
a two-point function or a sum rule); the other is the very high precision
of the comparison and also that it is possible to test form factors
of many particles which have never been tested using spectral sums,
mostly due to the fact that usually their contribution to spectral
expansions is extremely small, and evaluating it also involves calculating
multidimensional integrals to very high precision, which a numerically
difficult task. The second problem is actually related to the first,
since due to the smallness of the contribution from higher particle
terms all the lower ones must be evaluated to sufficiently high precision.
Our approach, in contrast, makes it possible to have a test of entire
one-dimensional sections of the form factor functions using the volume
as a parameter, and the number of available sections only depends
on our ability to identify multi-particle states in finite volume.

Our results can also be viewed in the context of finite volume form
factors \cite{finitevolFF} (which is also related to the problem
of finite temperature form factors; for a review on the latter see
\cite{Doyon} and references therein). The relations (\ref{eq:ffrelation},
\ref{eq:genffrelation}) give finite volume form factors expressed
with their infinite volume counterparts to all orders in $1/L$ (where
$L$ denotes the volume), i.e. up to exponentially decaying terms
in $L$. This gives finite volume form factors in large volume with
very high precision. On the other hand, what we determine numerically
in TCSA are actually the finite volume form factors themselves, which
is an approach that primarily works in small enough volume due to
the truncation errors. In the Lee-Yang case, the combination of the
two approaches gives the finite volume form factors involving up to
four particles with better than $10^{-3}$ relative precision, as
demonstrated by the excellent agreement in the scaling region of TCSA
where their domains of validity overlap. For the Ising model the numerical
precision is not as good, but with some care a precision of around
$10^{-3}$ can be achieved for most of the matrix elements considered
in this paper.

An open question which is not discussed in this paper is the case
of matrix elements with disconnected pieces. Results on such matrix
elements are already available, but we postpone them to a followup
paper \cite{crossing} where we also plan to discuss many theoretical
issues related to crossing and disconnected contributions in finite
volume.

\subsection*{Acknowledgments}

We wish to thank Z. Bajnok and L. Palla for useful discussions. This
research was partially supported by the Hungarian research funds OTKA
T043582, K60040 and TS044839. GT was also supported by a Bolyai J\'anos
research scholarship.

\end{document}